\documentclass[9pt,twocolumn,twoside]{pnas-new}

\usepackage[normalem]{ulem}
\usepackage{xcolor}
\usepackage{pbox}
\usepackage{url}

\newcommand\redsout{\bgroup\markoverwith{\textcolor{red}{\rule[0.5ex]{2pt}{0.4pt}}}\ULon}

\templatetype{pnasresearcharticle} 

\title{Local alliances and rivalries shape near-repeat terror activity of al-Qaeda, ISIS and insurgents}

\author[a,b]{Yao-Li Chuang}
\author[c]{Noam Ben-Asher} 
\author[a,b,1]{Maria R. D'Orsogna}

\affil[a]{Dept. of Biomathematics, UCLA, Los Angeles, CA 90095-1766}
\affil[b]{Dept. of Mathematics, CSUN, Los Angeles, CA 91330-8313}
\affil[c]{U.S. Army Research Laboratory Adelphi, MD, 20783}

\leadauthor{Chuang} 

\significancestatement{
We examine near-repeat activity patterns of al-Qaeda, ISIS and local insurgents, 
whereby a first terrorist attack temporarily increases the likelihood of a second 
one by the same group. We observe heightened near-repeat activity for all organizations 
in six geographic clusters, and quantify the effect to persist within 20km and 4 to 10 weeks 
after the first event. Near-reaction patterns, where two distinct groups react to 
each other's activities, depend on the adversarial, neutral or collaborative relationship
between parties at the local level. We find no evidence of outbidding, whereas
terrorist and state activities mutually reinforce one another.
Our results may be useful for counter-terrorism decision-making and strategic resource allocation; 
near-repeat patterns may offer insight into local power structures. 
}

\authorcontributions{All authors contributed equally to this work.}
\authordeclaration{We declare no conflict of interest}
\correspondingauthor{\textsuperscript{1}To whom correspondence should be addressed. E-mail: dorsogna@csun.edu}


\begin{abstract}
We study the spatiotemporal correlation of terrorist attacks by al-Qaeda,  
ISIS, and local insurgents, in six geographical areas identified via $k$-means clustering applied to
the Global Terrorism Database.  All surveyed organizations exhibit near-repeat activity whereby a prior attack increases the likelihood of a subsequent one
by the same group within 20km and on average 4 (al Qaeda) to 10 (ISIS) weeks. 
Near-response activity, whereby an attack by a given organization elicits further attacks from a different one, is found to
depend on the adversarial, neutral or collaborative relationship between the two.
When in conflict, local insurgents respond quickly to attacks by global terror groups 
while global terror groups delay their responses to local insurgents, leading to an
asymmetric dynamic. When neutral or allied, attacks by one group enhance the response likelihood of the other,
regardless of hierarchy. These trends arise consistently in all clusters for which data is available. 
Government intervention and spill-over effects are also discussed; we find no evidence of outbidding. 
Understanding the regional dynamics of terrorism may be greatly beneficial in 
policy-making and intervention design. 
\end{abstract}

\dates{This manuscript was compiled on \today}
\doi{\url{www.pnas.org/cgi/doi/10.1073/pnas.XXXXXXXXXX}}

\begin{document}

\maketitle
\thispagestyle{firststyle}
\ifthenelse{\boolean{shortarticle}}{\ifthenelse{\boolean{singlecolumn}}{\abscontentformatted}{\abscontent}}{}

\dropcap{T}errorist activities by al-Qaeda (AQ) and
the Islamic State of Iraq and Syria (ISIS) have brought violence and destruction to 
the Middle East and the world \cite{WAT16}. 
Many historical, political, religious 
motivations lie behind this unrest 
and several complementary perspectives have been offered to explain it 
\cite{HAS14,ZEL14,LAS16,ROU16}.
Recent advances in data collection have allowed for the thorough updating of terrorist databases; 
being able to extract information from them may help gain new insight and 
yield novel counterterrorism opportunities.

This work examines the spatiotemporal correlation of terrorist attacks perpetrated worldwide by AQ and ISIS,
focusing on the post-2014 era, when they began functioning as independent entities.  
On the local scale, the AQ/ISIS dynamics may be affected by independent militias or insurgents acting as 
rivals or allies to either. These groups are often entrenched in their communities and may be quite powerful, even eclipsing AQ or ISIS. 
Using near-repeat analysis \cite{JOH07,TOW08,JOH09,SHO09,SHO10,ORN17}
on data derived from the Global Terrorism Database (GTD), we analyze
patterns of attack for three classes of terror groups (AQ, ISIS and local militias) 
in six geographical clusters. The latter are identified through $k$-means clustering
without imposing any \textit{a priori} geographical constraints, such as national borders. 
The constituency of each class varies across clusters:
AQ or ISIS may have the most combatants in some, in others, local militias may be the most
numerous. Despite cluster heterogeneity, our findings indicate universal near-repeat activity, whereby an attack 
by a given group temporarily raises the probability of further attacks by the same one within 20km
over 4--10 weeks. The variability depends on location and organization, but 
groups with fewer combatants are always found to display the longest
period of enhanced near-repeat. 
While insurgents are present in all clusters, AQ and ISIS may or may not be;
where they do co-localize, one is 
numerically superior to the other and in conflict with insurgents. 
The two most numerous groups (insurgents and either
AQ or ISIS) are regarded as major players; the minor, numerically inferior one between AQ/ISIS,
aligns with either of the major two.
We also examine patterns of responding near-repeat, in short near-reaction, 
whereby an attack by a given group elicits a response from a different one. 
Insurgent activity intensifies after an attack by a major rival (either AQ/ISIS) while the latter 
suppresses its activity following insurgent attacks, leading to an asymmetric dynamic. 
Aligned groups, allies or neutral parties united against a common enemy, 
reinforce each other's activities so that regardless
of which one strikes first, the other will always intensify its attacks
within 20km. No evidence of outbidding is found, whereas terrorist
activity increases in response to government action in all surveyed
clusters. Our findings underline the importance of local geography and hierarchies in 
understanding terrorism.

\begin{figure*}[t]
\centering
\includegraphics[width= 1.0\linewidth]{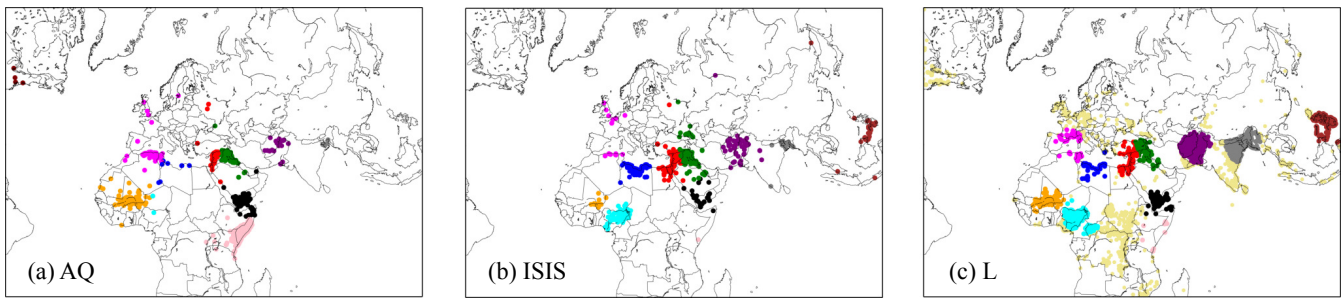}
\caption{Color-coded geographic clusters of post-2014 AQ/ISIS attacks.
Twelve clusters are identified via $k$-means clustering for their attacks
recorded in GTD. Each cluster is labeled by the country with the majority 
of AQ/ISIS attacks: Iraq, Somalia, Syria, Yemen,
Nigeria, Pakistan/Afghanistan, Libya, Algeria/E.U., Mali, Philippines,
Bangladesh, and the U.S. The above ordering follows the total number of AQ/ISIS attacks 
per cluster after the AQ-ISIS rift in 2014.  Only the first 6 contain enough data for 
near-activity analysis. Panel (a) displays AQ post-2014 activity in each cluster; 
panel (b) ISIS activity. Each dot represents one attack.
Panel (c) shows post-2014 attacks perpetrated by local insurgent groups 
co-localizing with these twelve clusters using the same color codes, whereas
attacks located outside the same twelve clusters are denoted by light yellow dots.
}
\label{FIG:CLUSTER}
\end{figure*}

 \section*{Historical background}
AQ was founded by Osama bin Laden during the Soviet-Afghan War in the 1980s. 
Due to intense international pressure after the 9/11 attacks, 
AQ evolved into a decentralized ``dune''-like organization
\cite{MIS05}: its affiliates were encouraged to operate 
independently while still being part of its broader network.  Among
the myriad of players, ISIS rapidly emerged as a strong
challenger to AQ's lead.  ISIS was originally established in Jordan
by Abu Musa al-Zarqawi in the late 1990s under the name Jamaat al-Tawhid wal-Jihad (JTJ). 
In 2004 al-Zarqawi pledged allegiance to AQ and renamed JTJ al-Qaeda in Iraq (AQI).
Instructed by AQ to establish a formal governing body, in early 2006 
AQI merged with other insurgent groups to form the 
Mujahideen Shura Council (MSC) and later
announced the creation of the Islamic State of Iraq (ISI), a centralized group
aiming to occupy the northern Iraqi territories by the Syrian border. 
As the Syrian civil war intensified in 2013, ISI became ISIS 
and expanded into Syria without consulting AQ's general command,
creating tension with Jabhat al-Nusra (JN), a Syrian 
AQ affiliate.  After repeated attempts to control its
expansion, AQ disavowed ISIS on Feb 2 2014. In the years since, AQ continued fostering local collaborations, 
including with al-Shabaab in Somalia and Boko Haram in Nigeria, 
while ISIS expanded beyond its Iraq/Syria base to control 
remote territories in Yemen, Afghanistan and Pakistan.
Given their more adversarial nature and efficient media usage, counter-terror efforts focused on constraining
ISIS, leading to concerns of indirectly strengthening AQ \cite{ROL11}.
Since imbalanced intervention may be ineffective \cite{GAO17}, 
better understanding the interplay among groups may lead to strategies
that do not inadvertently bolster any of them. 

\begin{table*}[t]
\centering

\caption{\label{TAB:CLUSTER_GROUPS} Active terrorist groups in six
significant clusters and their relative strength as measured by
estimated numbers of combatants
}

\begin{tabular}{|l|c|c|c|}
  \hline
  Clusters & Major 1 (number of combatants) & Major 2 (number of combatants) & Minor (number of combatants) \\
  \hline
  \hline
  Iraq & ISIS (200K$^1$) & & \pbox{5cm}{al-Naqshabandiya Army (50K$^2$), PKK (32K$^3$), Badr Brigades (10K -- 15K$^4$).} \\
  \hline
  Syria & ISIS (200K$^1$) & \pbox{5cm}{The Islamic Front (70K$^5$), Free Syrian Army (40K -- 50K$^1$), PKK (32K$^3$).} & 
  AQ:JN (20K$^6$) \\
  \hline
  Somalia & AQ:al-Shabaab (7K -- 9K$^7$) & & \\
  \hline
  Yemen & \pbox{5cm}{Houtis (100K$^8$), Southern Movement (N/A).} & AQAP (12K$^9$) & IS:Yemen Province (300$^{10}$)\\
  \hline
  Nigeria & IS:Boko Haram (20K$^{11}$) & & Fulani extremists (N/A) \\
  \hline
  \pbox{1cm}{Afghan-Pakistan} & Taliban (60K$^{12}$) & & IS:Khorasan Chapter (3K$^{13}$) \\
  \hline
\end{tabular}
\begin{justify}
\addtabletext{Estimated number of combatants are in parentheses; N/A indicates 
they are not available. Major 1-2 groups are known from journalistic reports to be in conflict.
Only the most prominent groups of the L class are listed; attacks
from smaller ones are also included in this study.  
References: Cockburn P., (2014) {\em The Independent}$^1$; Freeman C., (2013) {\em The Telegraph}$^2$;
Tarihi G., (2015) {\em Sabah}$^3$; George S., (2014) {\em Foreign Policy}$^4$; 
Hassan H., (2014) {\em Foreign Policy}$^5$; Kareem B., (2016) {\em OGN News}$^6$;
{\em BBC News} (2017)$^7$; Almasmari H., (2011) {\em CNN}$^8$; 
Roggio B., (2010) {\em FDD's Long War Journal}$^9$; 
Aboudi S., (2015) {\em Reuters}$^{10}$; {\em BBC News} (2015)$^{11}$; Dawi A., (2014) 
{\em Voice of America}$^{12}$; Seldin J., (2017) {\em Voice of America}$^{13}$.}
\end{justify}

\end{table*}

\section*{Materials and Methods}

We examine the terrorist activities of AQ, ISIS 
and local groups as listed by the GTD
which catalogued global terrorist attacks between Jan 1 1970 and Dec 31 2017
\cite{GTD18}.
We mostly focus on the post-2014 era, when AQ disavowed ISIS.

\begin{figure*}[h]
\centering
\includegraphics[width=1.0\linewidth]{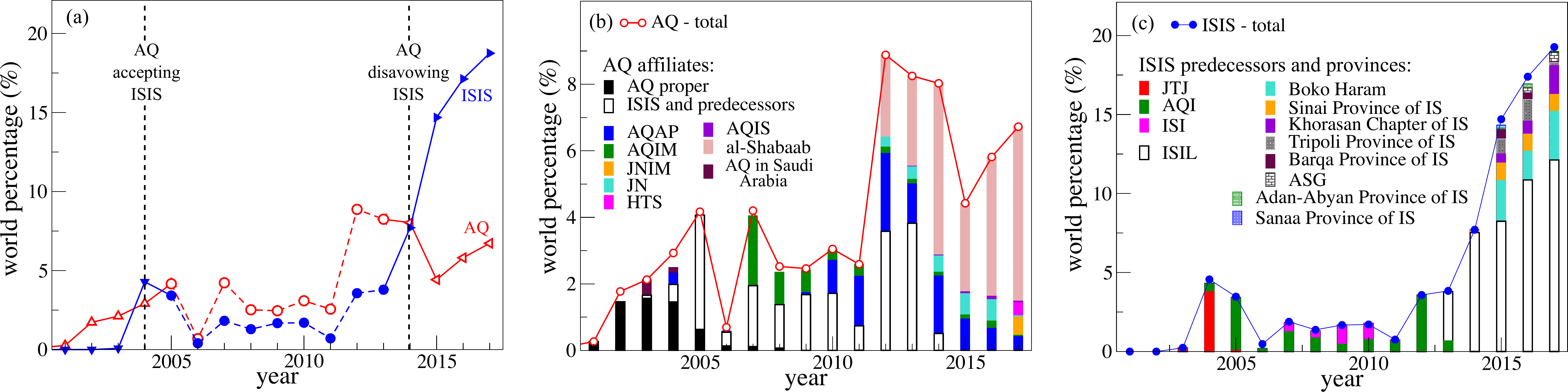}
\caption{(a)  AQ and ISIS terrorist activity
as percent of world total.
Between 2004-2014 the AQ curve includes contributions from ISIS and its precursors.
Pre-2004 data in the ISIS curve is due to JTJ, its first precursor;
post-2014 data are attacks by ISIS operating independently from
AQ. The limited activity in 2006 is most likely due to the great regional turmoil at
the time, with the media (and the GTD) unable to identify attackers.
Indeed, the total number of attacks in the Middle East increased
from 2005 to 2006 but so did the number
of incidents with unknown perpetrators.
(b) Yearly AQ affiliate attacks. (c)  Yearly ISIS affiliate attacks.
Note the diverging trends after the 2014 AQ/ISIS rift.}
\label{FIG:IYEAR}
\end{figure*}
\subsection*{Data selection}

The GTD lists a maximum of three confirmed or suspect perpetrator groups per attack. 
We discard entries with unknown offenders (46$\%$ of the total).
If at least one of the perpetrators is an AQ affiliate, the record is assigned to AQ; similarly for ISIS.  
If AQ and ISIS are responsible for the same attack, it is assigned to both. 
Finally, if neither is listed as the perpetrator, the record is assigned to the L class 
(local militias/insurgents). 
Affiliates were added to the AQ/ISIS classes only after their formal acceptance; 
since during 2004-2014 ISIS predecessors were recognized as AQ affiliates,  
their attacks within this period are assigned to both the AQ and ISIS classes.
Full affiliate AQ/ISIS lists are in the SI Appendix.

\subsection*{Cluster analysis}

To examine the spatiotemporal distribution of AQ/ISIS attacks
we first identify geographic areas where they co-localize using 
$k$-means clustering \cite{MAC67}.  The sole input here is attack location;
no other constraints, such as state borders, are used
(SI Appendix).
We identify twelve clusters and determine
the geographic centers and standard deviations (STD) of each. 
As shown in Figs.\,\ref{FIG:CLUSTER}(a)-(b) clusters are found to 
mostly coincide with geo-political boundaries, due to increased border security, 
attacks occurring in civic centers in the interior,  
or terrorist familiarity with terrain/culture. In some cases we observe spill-over or domino effects, due to weak borders or
historical/political precedents \cite{LAF18}. The Afghan-Pakistan cluster arises due to militant groups residing
between the two countries; the Syria cluster is found to include Lebanon due 
the Syrian civil war spilling over to its neighbor; the Nigeria cluster
includes small portions of neighboring Chad, Cameroon and Niger, as
Boko Haram attempted to evade government scrutiny; 
since AQ affiliated al-Shabaab often launched attacks from Somalia into neighboring Kenya the two form a 
unique cluster. Specific geographical conditions may thus allow terrorism to spread
across borders. To add an L class attack to one of the clusters identified above we calculate its distance from
the center of the nearest AQ/ISIS cluster; if the distance is within three STDs from its center, the L class attack is
assigned to this cluster, else it is discarded. 
The resulting distribution is shown in Fig\,\ref{FIG:CLUSTER}(c).  
The six clusters containing enough events to justify further analysis are
listed in Table\,\ref{TAB:CLUSTER_GROUPS}. We focus on them in the remainder of this work.
Major and minor groups are labeled on the basis of their
estimated number of combatants. The two major groups in Syria and Yemen are known rivals.

\subsection*{Near-repeat analysis}

\begin{figure*}[t]
\centering
\includegraphics[width=1.0\linewidth]{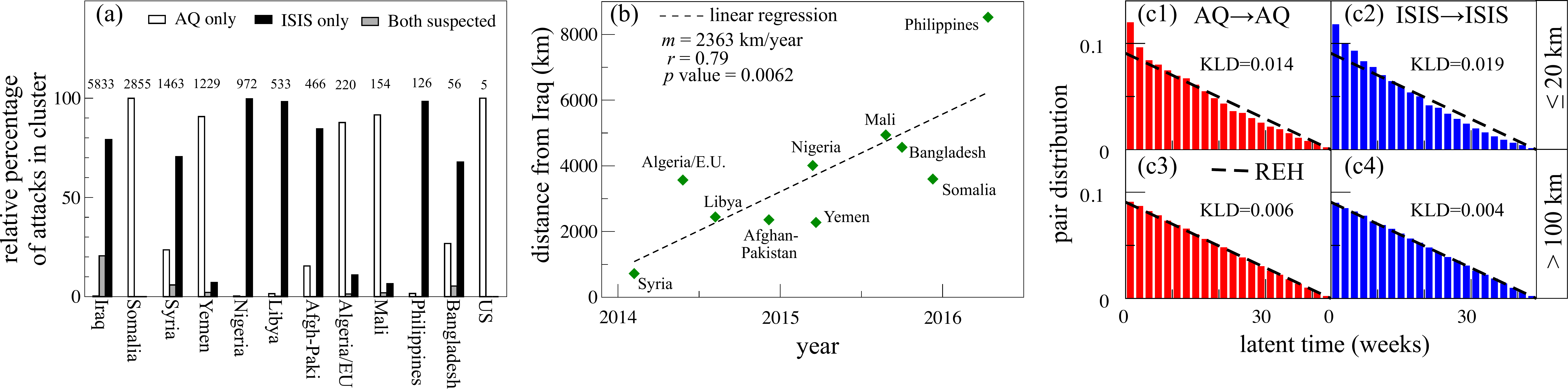}
\caption{(a) Attack fractions per cluster attributable to AQ and ISIS, 2001-2017.
(b) First ISIS attack dates per cluster
versus distance from Iraq post-2014.
A linear regression yields a spread of roughly $2400$km/year 
with correlation $r=0.79$ and a significant $p$ value of $0.006$.
(c) Near-repeat patterns for 
AQ$\rightarrow$AQ and ISIS$\rightarrow$ISIS
worldwide, 2014-2017. 
Each bar contains data binned over two weeks.
Deviations from the REH are estimated using KLD.
(c1) The distribution of AQ$\rightarrow$AQ attacks separated by less than 
20km deviate from the REH over the first 4 weeks, indicating increased near-repeat 
likelihood. (c2) For ISIS$\rightarrow$ISIS attacks the enhanced 
repeat likelihood is over the first 10 weeks. (c3) and (c4) AQ$\rightarrow$AQ and
ISIS$\rightarrow$ISIS attacks separated by more than 100km follow the 
REH, suggesting negligible correlation between repeat events.}
\label{FIG:COMBINE}
\end{figure*}

Terrorist events within each cluster are examined using near-repeat analysis tools developed in criminology
\cite{TOW08,JOH09,SHO09,SHO10,ORN17}.
Near-repetition within the ``fixed window'' method  is quantified by comparing the distribution of 
given data to a hypothetical distribution of random, independent events  \cite{SHO09, SHO10}. 
This is done by partitioning the given timespan into several windows of time $w$; 
within each window the time difference $0 < t < w$ and the spatial distance
between every pair of events are calculated.
The time distribution for events occurring within a
a maximum distance $d$ are subsequently compiled from all windows. 
This data-derived distribution is then compared to the random-event hypothesis (REH) 
\begin{equation}
   P_d (t) = 2 \frac{w-t}{w (w - 1)}.
   \label{EQ:REH}
\end{equation}

\noindent
Eq.\,\ref{EQ:REH} is the random probability distribution for finding a pair of events separated by a time interval $t$ within $w$, assuming events are uniformly distributed. 
For example, if $w=7$ days, the probability of two events being separated by $t=1$ day is proportional to 
the number of pairs separated by one day in one week: 
Mon-Tue, Tue-Wed through Sat-Sun, for a total of six possibilities. Conversely, there is only one possible
pair of events separated by $t=6$ days, Mon-Sun. $P_d(t)$ is thus proportional to $w - t$; 
the normalization prefactor $2/w (w-1)$ in Eq.\,\ref{EQ:REH} ensures that 
$\sum_{t=1}^{w} P_d(t) = 1$.  Deviations from $P_d (t)$ indicate a non-random likelihood
for an event to repeat after a prescribed time $t$.
Contrasting observed distributions to $P_d (t)$ is equivalent to performing 
a Knox ratio analysis \cite{KNO64,BRA12,ORN17}.  
However, the latter requires 
sampling and Monte Carlo simulations to generate a randomized distribution for data comparison. 
Eq.\,\ref{EQ:REH} requires fewer assumptions and eliminates the need for simulations.
We set $w=44$ weeks, 
guaranteeing all near-repeat effects are captured. 
Feb 2 2014, the official AQ/ISIS rift date,
is used as a starting point from which successive windows of $w$ periods
are generated. 
To verify whether any biases were introduced, 
different $w$ values and start dates were used; results 
remained essentially unchanged.  Same-day attacks are discarded, as it is not possible to determine whether
they are part of a coordinated campaign or independent events \cite{DEL13}.

\subsection*{Government intervention}

Since government (G) activity is not part of the GTD, 
we utilize the Uppsala Conflict Data Program (UCDP) dyadic dataset \cite{SUN13}
to obtain lists of state-sponsored counterterrorism events.
These are then cross-listed with terrorist attacks from the GTD 
to study terrorist/counterterrorist interplay. 
Sufficient data exists only for Iraq, Somalia and the Afghan-Pakistan clusters
where government activities were targeted at ISIS, AQ and the Taliban, respectively.
We use post-2014 data for Iraq, post-2012 data for Somalia, when al-Shabaab joined AQ and the
Federal Government of Somalia was established, and
post-2003 data for the Afghan-Pakistan cluster, when the Taliban launched large scale insurgencies
against the Afghan government.
There is not enough UCDP data for Nigeria or Syria 
\cite{HOG19}; it is not possible to identify
a legitimate government in Yemen due to the ongoing civil war.

\section*{Results}

\subsection*{Al-Qaeda and ISIS activities}
Terrorist attacks executed by AQ and ISIS between 2001-2017 are shown in 
Fig.\,\ref{FIG:IYEAR}(a). Until 2011, AQ contributed to less than $5 \%$ of global activity, 
the ISIS precursors even less.
In 2012, AQ's activities increased to roughly $10 \%$ of world total, mostly fueled by 
ISIS and al-Shabaab as shown in Fig.\,\ref{FIG:IYEAR}(b).
After the 2014 AQ/ISIS rift the percent of attacks by AQ proper stagnated, 
while ISIS's activity grew rapidly, surpassing AQ to reach almost 
$20\%$ of world total in 2017.  Fig.\,\ref{FIG:IYEAR}(c) shows that this increase is due to greater activity of ISIS proper, 
but also to the emergence of affiliate terrorist groups. 
AQ's activities post-2014 are mostly due to al-Shabaab in Somalia.
The distribution of terrorist attacks within the twelve geographical 
clusters identified in Fig.\,\ref{FIG:CLUSTER} is illustrated in 
Fig.\,\,\ref{FIG:COMBINE}(a). 
In Iraq, AQ attacks mostly cease after the 2014 AQ/ISIS rift while concurrent AQ/ISIS activities persist 
in Syria and Yemen as also shown in the time-resolved plot in 
SI Appendix Fig.\,S2. 
AQ is mostly absent in Nigeria, Libya, and the Philippines; 
ISIS is inactive in Somalia and in the U.S. 
We also observe ISIS spreading from its Iraq base 
to neighboring countries through the terror ``wave''  in Fig.\,\ref{FIG:COMBINE}(b). 
Here, the date of ISIS's first attack in each cluster is plotted as a function of its distance from Iraq. 
The progressive spread of ISIS activities also emerges from 
SI Appendix Fig.\,S3,  where attacks are geo-spatially mapped over time. These findings
are consistent with ISIS's focus on expansion while AQ's activities vary in a more
random manner, consistent with its decentralized ``dune''-like structure.

\begin{figure*}[t]
\centering
\includegraphics[width=1.0\linewidth]{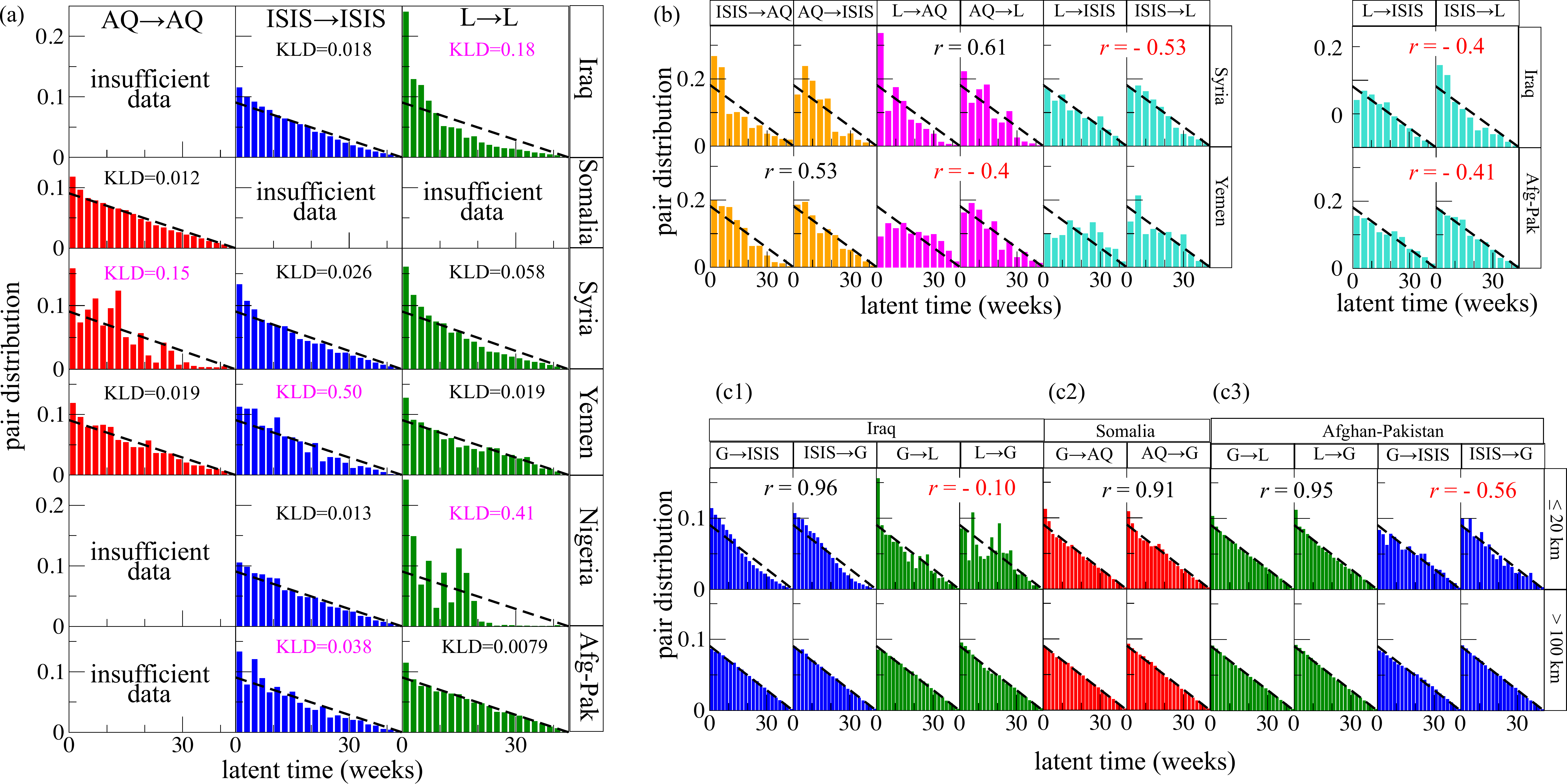}
\caption{
(a) Near-repeat patterns 
AQ$\rightarrow$AQ, ISIS$\rightarrow$ISIS, and L$\rightarrow$L, 
in six clusters post-2014. Each bar contains data binned over 2 weeks; 
all pairs of events are separated by less than 20km.
Heightened near-repeat likelihood is observed at least over the first 4 weeks
in all panels. Near-repeat tendencies are more prominent for
the minor classes, 
as indicated by their KLD values shown in magenta. (b) Near-reaction patterns A$\to$B, B $\to$A, where
$\{$A,B$\}$ = $\{$AQ, ISIS, L$\}$ in clusters where their post-2014 activities overlap. 
The first class in each panel is the first to attack; 
the second responds within 20km after the given latent time.
Each bar contains data binned over 4 weeks.
The correlation coefficient $r$ between mirror panels is included
for moderate correlation ($|r| > 0.3$). All known major rivals within each cluster
show negative correlation. Panels showing
positive correlation are either allies (AQ and L, Syria)
or share ideology (AQ and ISIS, Yemen). 
(c1)-(c3) Terrorist/government (G) near-reaction patterns;  in each cluster the correlation coefficient $r$ between G and the main class
(ISIS in Iraq, AQ in Somalia and L in the Afghan-Pakistan cluster) is shown in black. Each bar contains data binned over 2 weeks.
Large $r$
values suggest strong interplay. (c1) In Iraq, ISIS intensifies its activity compared to the REH after G
intervention over 4 months; G increases its activity over 4.5 
months after ISIS attacks. The agile L class responds to the same
G operations by intensifying its attacks over 2 weeks. No consistent patterns arise in L $\to$ G as can be expected.
(c2) In Somalia similar trends arise with enhanced AQ activity after G intervention (4 weeks); G also increases its activity in response to AQ (4 weeks).
(c3) In the Afghan-Pakistan cluster near-reaction is enhanced in the G $\to$ L (4 weeks) 
and L $\to$ G (6 weeks) panels where L is the Taliban.  Although data is limited, 
we observe a delayed response from ISIS and an asymmetric
G $\to$ ISIS and ISIS $\to$ G dynamic with $r = -0.56$, 
typical of near-repeat patterns involving transnational groups. 
Events separated by more than 100km follow the REH. 
Panels with less than 100 data points are not analyzed due to insufficient data.
The L classes are as listed in Table\,\ref{TAB:CLUSTER_GROUPS}.}
\label{FIG:COMBINE2}
\end{figure*}

\subsection*{Near-repeat activity}

Near-repeat results are shown in Fig.\,\ref{FIG:COMBINE}(c) where we plot the latent time $t$ between every pair of 
AQ after AQ (i.e., AQ$\rightarrow$AQ) or ISIS after ISIS (i.e., ISIS$\rightarrow$ISIS) 
attacks separated by less than 20km or more than 100km, worldwide and 
post-2014. For comparison the REH distribution in Eq.\,\ref{EQ:REH} is also shown. 
We use the Kullback-Leibler divergence (KLD, SI Appendix) 
to quantify the deviation of attack distributions from the REH. 
Fig.\,\ref{FIG:COMBINE}(c1) shows that during the first 4 weeks after an AQ attack, the probability for 
a near-repeat attack within 20km is elevated with respect to the REH. 
ISIS exhibits an even longer period of elevated near-repeat probability within 20km: 10 weeks,
as shown in Fig.\,\ref{FIG:COMBINE}(c2).  At distances beyond 100km, near-repeat effects are negligible and the 
latent time distributions converge to the REH, as seen in Figs.\,\ref{FIG:COMBINE}(c3)-(c4). 
Henceforth, we use 20km as the maximal distance for near-repeat events.
Only six of the twelve clusters identified in Fig.\,\ref{FIG:CLUSTER}
contain enough data for near-repeat analysis.  Fig.\,\ref{FIG:COMBINE2}(a)
shows the resulting patterns after the 2014 AQ/ISIS rift.  
All panels show elevated near-repeat probability for at least 4 weeks. This period 
is extended for the minor group in a cluster, as identified in Table\,\ref{TAB:CLUSTER_GROUPS},
which displays larger deviations from the REH, 
larger KLD values, and longer near-repeat periods compared to the major classes.
The trend is consistent across clusters. For example, ISIS in Iraq and Syria, L (Taliban) 
in the Afghan-Pakistan region, and L (Houthi and Southern Movement ethnopolitical
rebels) in Yemen, AQ (al-Shabaab) in Somalia are all major players, and display 
enhanced repeat likelihood over 4 weeks. For ISIS (Boko Haram) in Nigeria the period is 6 weeks. 
In clusters where minor players are also present, they 
always display longer periods of enhanced near-repeat
compared to the major groups listed above. The longest period is 14 weeks for ISIS, the minor player in Yemen.

\subsection*{Near-reaction activity}

To analyze the post-2014 interplay among AQ, ISIS and the L groups in each cluster
we consider the latent time distribution 
for A$\rightarrow$B events where a class A attack is followed by a
class B attack within 20km for $\{$A,B$\} = {\{\mbox{AQ, ISIS, L}}\}$
and A $\neq$ B.  We do not expect symmetry between the A$\rightarrow$B and B$\rightarrow$A panels
as power structures may be imbalanced.  We omit Somalia, 
with no significant ISIS or L presence, and Nigeria
where attacks by ISIS (Boko Haram) and L-class Fulani extremists display negligible 
overlap within $20$km.  
Fig.\,\ref{FIG:COMBINE2}(b)  shows data deviating significantly 
from the REH in all clusters indicating enhanced or hindered near-reaction. 
To interpret these results and determine whether the order in which the A,B classes strike and respond
is relevant, we calculate the Shannon entropy contribution $E_i$ for each of the 
A$\rightarrow$B and B$\rightarrow$A datasets in Fig.\,\ref{FIG:COMBINE2}(b). 
The $E_i \equiv \hat p_i \ln (\hat p_i / p_i)$ terms allow to compare datapoints 
$(t_i, \hat p_i)$  from Fig.\,\ref{FIG:COMBINE2}(b) with the corresponding REH $(t_i, p_i)$ where
$p_i = P_d (t_i)$ as evaluated from Eq.\,\ref{EQ:REH}. Finally, 
the correlation coefficient $r$ between the two derived datasets 
$\{ E_i^{A \rightarrow B}\}$ and $\{ E_i^{B \rightarrow A}\}$ is calculated.
We find that within each cluster, only one of the A$\rightarrow$B, B$\rightarrow$A 
panels is marked by appreciable negative correlation ($r < -0.3$, in red).
The respective $\{\rm{A,B}\}$ classes are: in Iraq, $\{\rm{A,B}\}$ = $\{$ISIS, L$\}$; in Syria,
$\{\rm{A,B}\}$ = $\{$ISIS, L$\}$; in Yemen, $\{\rm{A,B}\}$ = $\{$AQ, L$\}$; 
in Afghan-Pakistan  $\{\rm{A,B}\}$ = $\{$ISIS, L$\}$. The two parties involved are
always either AQ/ISIS (global organizations) and the local L-class, and are
always known to be in conflict. 
Table\,\ref{TAB:CLUSTER_GROUPS} indicates that in all cases 
the two above classes are also the largest in their clusters. 
Our finding of a negative correlation suggests that  an L-class group responds promptly to an attack by AQ/ISIS 
while the latter's response is delayed.  If a third class is present, it is always the minor player
listed in Table\,\ref{TAB:CLUSTER_GROUPS} and either collaborates (such as AQ and L-class rebels in Syria)
or shares a similar ideology (such as AQ and ISIS in Yemen)
with either of the two major ones. Fig.\,\ref{FIG:COMBINE2}(b) shows that the minor party always correlates positively with the 
major player it aligns with.  Correlations between the minor group and the other major player
are not significant indicating that the major player neglects its smaller adversary to focus on its main rival.
These findings are consistent across surveyed clusters, 
and emerge for pre-2014 data as well 
(SI Appendix).

\subsection*{Government intervention}

We examine terrorist/state interplay in the Iraq, Somalia and Afghan-Pakistan clusters
in Figs.\,\ref{FIG:COMBINE2}(c1)-(c3). In Iraq, the post-2014 G (UCDP data) $\to$ ISIS
(GTD data) and ISIS $\to$ G near-reaction panels show both parties displaying enhanced near-response 
within 20km over several months. There is no data for government action explicitly aimed at the L-class in Iraq, 
however we can examine whether government operations directed at 
ISIS indirectly affect L-class activities. Thus, in Fig.\,\ref{FIG:COMBINE2}(c1) we also study
G $\to$ L (GTD data) near-reaction activity. L responses spike for 2 weeks after G action,
most likely as a short-lived immediate reaction to any nearby event,
regardless of intended targets. No clear L $\to$ G trends emerge, implying that government activity aimed
at ISIS is not influenced by prior L-class activities.
Similar trends are observed for the G $\to$ AQ and AQ $\to$ G post-2012 
panels in Fig.\,\ref{FIG:COMBINE2}(c2) for Somalia, 
where no L-class group is present, and in the G $\to$ L and L $\to$ G post-2003 panels in Fig.\,\ref{FIG:COMBINE2}(c3) 
for the Afghan-Pakistan cluster (L here is the Taliban). 
We also include near-reaction between the small ISIS group present in the Afghan-Pakistan cluster
and government interventions aimed at the Taliban. 
Despite the small sample size of the ISIS attacks, the asymmetry between 
the G $\to$ ISIS and ISIS $\to$ G panels
in Fig.\,\ref{FIG:COMBINE2}(c3), is in agreement with earlier observations of 
delayed responses by global groups. 
Large correlation values in all clusters indicate strong interplay between 
terrorist/state actors, each aiming to signal their own supremacy. 
Except for Somalia, state operations persist slightly longer than terrorist activity.

\section*{Discussion}


Near-repeat patterns, often observed in 
urban crime,  also emerge for terrorist attacks by AQ, ISIS and mostly all other 
contemporary insurgent groups, 
despite heterogeneous
local conditions. The corresponding spatiotemporal scales (20km and 
4-10 weeks) are larger than for urban crime, 
where effects persist over a few hundred meters for
2-6 weeks 
\cite{JOH07, SHO09,SHO10}. 
This may be due to terror attacks being more impactful 
in terms of damage, media coverage, psychological effects;
longer times may also be required to orchestrate repeat
events.
Spatial clustering reveals territorial effects.  For example, AQ displays a 
longer period of elevated near-repeat than ISIS 
in Syria, but the trend is reversed in Yemen. In both cases, 
the minor organization exhibits more prominent near-repeat activity
than the major one, a consistent finding in this work.
Why? We note that numerically inferior groups may need to 
act more frequently to maintain visibility, especially in the presence of 
more established organizations. This is in agreement with
game-theoretic studies whereby sustained
violence allows small groups of radicals to stay relevant and grow
  \cite{SHO17}. Near-reaction patterns between major rivals
in the same cluster show asymmetric behavior. Typically, one of the two 
global terrorist organizations (AQ or ISIS) is in conflict with 
local militias/insurgents which respond promptly to 
AQ or ISIS attacks. Conversely, 
the response of the global terror group is delayed in time. This asymmetric behavior may be due 
to local militias having the agility and need to respond quickly,  
whereas global terror groups may require longer 
decisional times. The remaining group, typically with the least number of combatants, tends to align itself with one of
the major ones, executing copy-cat or supportive attacks. These aligned parties may be engaged in a leader-follower relationship,
the minor party may be a supportive ally of the major one, or the two may be neutral, non-hostile entities
united by the same ideology or intent against the same rival.
The trend is consistent with social balance theory which posits that
in a triadic relationship where two entities are already 
in conflict, the only balanced position the third can adopt is to 
align itself with one and oppose the other \cite{HEI46}.  
Recent studies on urban gangs confirm similar dynamics \cite{NAK19}. 
This triadic pattern is manifest in Syria, where post-2014, 
AQ affiliate JN began supporting the L-class antigovernment rebels, leading to ISIS$\rightarrow$L
rivalry/negative (r/n) correlation, AQ $\rightarrow$L aligned/positive (a/p)
correlation, and weak AQ $\rightarrow$ISIS correlation. 
In Yemen, AQ affiliate AQAP has been a longtime opponent to local insurgents
whereas ISIS established its provinces here in 2015.  AQ and ISIS are rivals but both regard
insurgent groups as their primary enemies, 
leading to AQ $\rightarrow$L r/n correlation, AQ $\rightarrow$ISIS a/p correlation
and weak ISIS $\rightarrow$L correlation. In the Afghan-Pakistan region, AQ has mainly assisted the Taliban; 
as ISIS attempted to gain territory, it engaged in open conflict with the Taliban.
We thus observe ISIS$\rightarrow$L (Taliban)  
and ISIS $\rightarrow$AQ 
r/n correlation. There is not enough data 
to compare AQ and the Taliban. In Iraq, there is no AQ presence post-2014, 
so we only observe ISIS$\rightarrow$L (ethnoreligious militias)
r/n correlation.  We find no 
evidence of outbidding \cite{BLO05} in the form of escalating responses 
under provocation or when casualties are involved 
(SI Appendix) consistent with existing studies 
 \cite{FIN12, HAM17}. 
State and terrorist action
are strongly correlated, 
with long term near-reaction deviations from the REH
(SI Appendix).
This indicates
both parties prioritize overcoming the other, and questions the immediate efficacy of 
state-sponsored military action.
Interestingly the G $\to$ ISIS panel in Iraq is the only one where ISIS does
not show a delayed response, suggesting that when priorities or clear enemies
arise, ISIS does have a fast reaction capability.

\section*{Conclusion}

We studied the spatiotemporal patterns of terrorist attacks by AQ, ISIS and local 
militias/insurgents by applying data-driven, unsupervised $k$-means clustering to the GTD. 
While near-repeat/reaction patterns
are observed in all clusters 
 \cite{LEW12, WHI13,TEN16}, 
the accompanying variations 
highlight the territorial aspects of terrorism \cite{BRA07} and the role played by local hierarchies,
even when global terrorist groups are present.  Near-repeat duration, 
inter-group dynamics and government response
depend more on a group's relative size and on its local relation to other groups 
than on whether or not it is part of a transnational organization. 
Understanding the local aspects of terrorism 
may help policy-makers better plan timing, permanence, and expectations of anti-terrorism 
intervention. 
Our correlative observations can help shape statistically optimal resource allocation
and logistics to more efficiently respond to future attacks.
Possible extensions could include 
constructing poly-order chains of events to distinguish attack methods and weapon/target types 
to further refine the understanding of repeat activity and the design of response protocols
\cite{BEH12, HAU10}.

\showmatmethods{} 

\acknow{
We acknowledge support from the Army Research Laboratory through grant ARL W911NF-16-2-0113 (to N.B.-A.), from the Army Research Office through grant ARO W911NF-16-1-0165 (to M.R.D), and from the National Science Foundation through grant DMS-1814090 (to M.R.D.). We thank V. Asal, A. Moghadam, and E. Miller for helpful discus- sions and the reviewers for valuable feedback. The views and conclusions contained in this document are those of the authors and should not be inter- preted as representing the official policies, either expressed or implied, of the Army Research Laboratory or the US Government. The US Government is authorized to reproduce and distribute reprints for government purposes notwithstanding any copyright notation herein.
}

\showacknow{} 
\bibliography{aqis} 

\end{document}



\maketitle

\SItext


\section*{AQ and ISIS affiliates, and L-class groups in the GTD}

\begin{table}[t]
\centering
\begin{tabular}{|c|l|}
  \hline
  1988 & Osama bin Laden founds AQ \\
  \hline
  1999 & Abu Musab al-Zarqawi establishes JTJ \\
  \hline
  2001 & JTJ moves its base to Iraq \\
  \hline
  Oct 17,  2004 & JTJ is renamed AQI and joins AQ \\
  \hline
  Jan 15, 2006 & AQI and other five groups form MSC\\
  \hline
  Oct 15, 2006 & ISI is officially established \\
  \hline
  Apr 8, 2013 & ISI is renamed ISIS and expands into Syria \\
  \hline
  Feb 2, 2014 & AQ formally disavows ISIS \\
  \hline
\end{tabular}
\caption{\label{TAB:TIMELINE} Timeline of relations between AQ and ISIS.}
\end{table}

Table \ref{TAB:TIMELINE} summarizes key events in the history of al-Qaeda (AQ) and ISIS, 
as described in the main text.  Our attack data is taken from the Global Terrorist Database (GTD)  available from the National Consortium for the Study
of Terrorism and Responses to Terrorism (START) which lists events between Jan 1 1970 and Dec 31 2017.
According to the GTD codebook, an event must meet two of the following three criteria to qualify as a terrorist attack: 
(1) it must have political, religious, or socioeconomic goals; 
(2) its intent must be to intimidate or coerce an audience larger than the immediate victims; 
(3) it must fall outside legitimate warfare activities, for example by deliberately targeting civilians.
Conventional military actions are not included.
Each entry contains time and location of the attack, name(s) of the perpetrator group(s), 
number of victims, weapon/target types, among others.

\subsection*{AQ and ISIS affiliates}

The first AQ entry recorded in the GTD is dated 1992
and took place in Yemen; the first ISIS attack is dated 2003, and was carried out by its 
predecessor Jamaat al-Tawhid wal-Jihad (JTJ) in Iraq.  Several challenges arise in identifying AQ and ISIS affiliates. 
First, they may have used various names and/or spellings depending on 
geopolitical context. Spellings may also depend on the sources
reporting terrorist activities. For example al-Qaeda is also translated as al-Qaida, al-Qa'ida, or el-Qaida, and ISIS
at times is referred to as ISIL (Islamic State of Iraq and the Levant). 
The GTD unified the English translation of group names, listing al-Qaeda and all other variants as al-Qaida, 
and ISIS and all other variants as ISIL. In our work we use the GTD classification but the more
common versions al-Qaeda (AQ) and ISIS.   Second, these global terrorist groups may
have incorporated (rejected) local groups into (from) their networks at given points in time, 
as dictated by circumstances; some associations may not even be fully clear due to the 
covert nature of terrorist activities.  
In our work, only groups that were officially accepted by AQ or ISIS 
into their networks, and for which verifiable documentation 
is available, are listed as affiliates. Furthermore, we list them only for the
period during which such recognition was granted. 
For example al-Shabaab of Somalia or Boko Haram of Nigeria joined AQ and ISIS, respectively,
after their founding, are included as AQ or ISIS affiliates after their official pledge dates.
Other groups that never formally pledged allegiance to either AQ or ISIS are assigned 
to the L class, such as the Taliban in the Afghan-Pakistan area. Finally, 
since ISIS predecessors were recognized as AQ affiliates between 2004 and 2014, 
their attacks within this period are assigned to both the AQ and ISIS classes.
Between 2001-2017, 6130 (9183) attacks are associated with AQ (ISIS), 
3383 (7878) of them took place post-2014.  
A total of 1328 joint AQ/ISIS attacks are listed between 2001-2017, of which only 72 are post-2014. 

The AQ class thus includes 
AQ proper (59 attacks in the GTD),
al-Qaeda in the Arabian Peninsula (AQAP, 1046 attacks), 
al-Qaeda in the Islamic Maghreb (AQIM, 282 attacks), 
al-Qaeda in the Indian Subcontinent (AQIS, 33 attacks), 
al-Qaeda in Yemen (12 attacks), al-Qaeda in Saudi Arabia (8 attacks),
Islambouli Brigades of al-Qaeda (5 attacks), Jadid al-Qaeda Bangladesh (JAQB, 3 attacks),
al-Qaeda Kurdish Battalions (AQKB, 2 attacks), 
al-Qaeda Network for Southwestern Khulna Division (2 attacks),
al-Qaeda in Lebanon (1 attack),
al-Qaeda Organization for Jihad in Sweden (1 attack). 
Groups founded under AQ's official oversight are also included, such as
Jabhat al-Nusra (JN, 344 attacks) and its successor Hay'at Tahrir al-Sham (HTS, 43 attacks) in Syria,
Jamaat Nusrat al-Islam wal Muslimin (JNIM, 59 attacks) in Mali. Al-Shabaab of Somalia formally
merged with AQ in 2012, and their 2947 attacks from that date onward were also added to
the AQ class.
Between 2003 and 2014, ISIS (499 attacks) and its predecessors
al-Qaeda in Iraq (AQI, 639 attacks), Mujahedeen Shura Council (MSC, 8 attacks),
and Islamic State of Iraq (ISI, 147 attacks) were part of the AQ network, so we include
their attacks in the AQ class during this period.
The Taliban of Afghanistan is considered an independent organization rather than
an AQ affiliate since, despite its well-known ties with AQ, it never formally pledged 
allegiance to AQ. Note that some of the events listed above
are associated with multiple AQ affiliates, either as perpetrators or suspects,
as a result the sum of the total number of attacks (obtained by summing the
numbers in parenthesis above) slightly exceeds the 6130 total. 

The ISIS class consists of ISIS proper
(5676 attacks), its predecessors, and its various remote provinces or chapters.
ISIS predecessors include
JTJ (1999 -- 2004, 47 attacks),
AQI (2004 -- 2013, 639 attacks), MSC (2006 -- 2006, 8 attacks), 
and ISI (2006 -- 2013, 147 attacks).
Beginning in 2013, ISIS established itself outside of
Iraq and Syria, through the Sinai Province (447 attacks) in Egypt, 
the Khorasan Chapter (396 attacks) in Afghan-Pakistan, 
the Tripoli Province (351 attacks), the Barqa Province (161 attacks),
the Fezzan Province (10 attacks) in Libya,
the Adan-Abyan Province (50 attacks), the Sanaa Province (30 attacks),
the Hadramawt Province (16 attacks), the Lahij Province (4 attacks),
the al-Bayda Province (2 attacks), the Shabwah Province (1 attack) in Yemen,
the Caucasus Province (17 attacks) in Russia,
the Algeria Province (10 attacks) in Algeria, 
the Najd Province (9 attacks) in Saudi Arabia,
the Bahrain Province (2 attacks) in Bahrain,
and a few other smaller branches in Egypt (22 attacks),
Bangladesh (38 attacks), the Greater Sahara (ISGS, 11 attacks),
and Jerusalem (10 attacks).
Boko Haram of Nigeria and Abu Sayyaf Group (ASG) of the Philippines,
respectively, joined ISIS in 2015 and 2016, and their respective 975 and 106 attacks 
from those dates onward are included in the ISIS class. 

\subsection*{L-class local militias/insurgent groups}

To analyze the interplay between AQ, ISIS and local militias/insurgent groups that define the L class,
we identified areas where attacks from the L class co-localize with the twelve AQ/ISIS clusters identified above.
We identify $38220$ worldwide L-class attacks between 2001-2017, of which $19116$ 
are post-2014. Of these, 13551 co-localize with the AQ/ISIS clusters. 
Here we list the names of local groups which contributed more than $20$
attacks in the respective clusters post-2014:

\begin{itemize}
\item {\bf Iraq cluster:} Kurdistan Workers' Party (PKK, 836 attacks, mostly in Turkey), Asa'ib Ahl al-Haqq (56 attacks),
Badr Brigades (54 attacks), Al-Naqshabandiya Army (26 attacks);
\item {\bf Syria cluster:} Hamas (Islamic Resistance Movement, 102 attacks), Free Syrian Army (92 attacks),
PKK (59 attacks), Ansar Bayt al-Maqdis (Ansar Jerusalem, 42 attacks), Ajnad Misr (32 attacks), Muslim Brotherhood (32 attacks),
Islamic Front of Syria (32 attacks), Jaysh al-Islam of Syria (30 attacks), Southern Front (28 attacks),
Palestinian Islamic Jihad (PIJ, 22 attacks);
\item {\bf Yemen cluster:} Houthi extremists (Ansar Allah, 975 attacks), Tribesmen (56 attacks),
Popular Resistance Committees of Yemen (25 attacks), Southern Mobility Movement of Yemen (21 attacks);
\item {\bf Nigeria cluster:} Boko Haram	(before joining ISIS, 615 attacks), Fulani extremists (429 attacks);
\item {\bf Afghan-Pakistan cluster:} Taliban (4128 attacks), Tehrik-i-Taliban Pakistan (TTP, 401 attacks),
Baloch Republican Army (BRA, 147 attacks), Hizbul Mujahideen (HM, 109 attacks), Lashkar-e-Taiba (LeT, 101 attacks),
Baloch Liberation Army (BLA, 68 attacks), United Baloch Army (UBA, 58 attacks), Lashkar-e-Islam of Pakistan (47 attacks),
Baloch Liberation Front (BLF, 42 attacks), Jaish-e-Mohammad (JeM, 42 attacks), Lashkar-e-Jhangvi (42 attacks),
Haqqani Network (31 attacks), Halqa-e-Mehsud (24 attacks);
\item {\bf Libya:} Ansar al-Sharia of Libya (55 attacks), Haftar Militia (25 attacks);
\item {\bf Mali cluster:} Ansar al-Dine of Mali (48 attacks), Macina Liberation Front (FLM, 28 attacks),
Movement for Oneness and Jihad in West Africa (MUJAO, 24 attacks);
\item {\bf Philippines cluster:} New People's Army (NPA, 1028 attacks), 
Bangsamoro Islamic Freedom Movement (BIFM, 296 attacks), Abu Sayyaf Group (ASG, before joining ISIS, 175 attacks),
Maute Group	(34 attacks);
\item {\bf Bangladesh cluster:} Communist Party of India - Maoist (CPI-Maoist, 204 attacks),
Pro Hartal Activists (91 attacks), Garo National Liberation Army (77 attacks), 
United Liberation Front of Assam (ULFA, 72 attacks), Gorkha Janmukti Morcha (GJM, 70 attacks),
Communist Party of Nepal - Maoist (CPN-Maoist-Chand, 66 attacks), 
National Democratic Front of Bodoland (NDFB, 53 attacks), 
National Socialist Council of Nagaland-Isak-Muivah (NSCN-IM, 51 attacks),
National Socialist Council of Nagaland-Khaplang (NSCN-K, 48 attacks),
Arakan Rohingya Salvation Army (ARSA, 46 attacks),
People's Liberation Front of India (44 attacks), People's Liberation Army of India (40 attacks),
Jamaat-E-Islami of Bangladesh (39 attacks), Bangladesh Nationalist Party (BNP, 23 attacks),
Achik Songna An'pachakgipa Kotok (ASAK, 20 attacks), Rohingya extremists (20 attacks);
\item {\bf Somalia, U. S., Algeria/E. U. clusters:} N/A.
\end{itemize}

\section*{$k$-means clustering analysis}

\begin{figure}[t]
\centering
\includegraphics[width=0.55\textwidth]{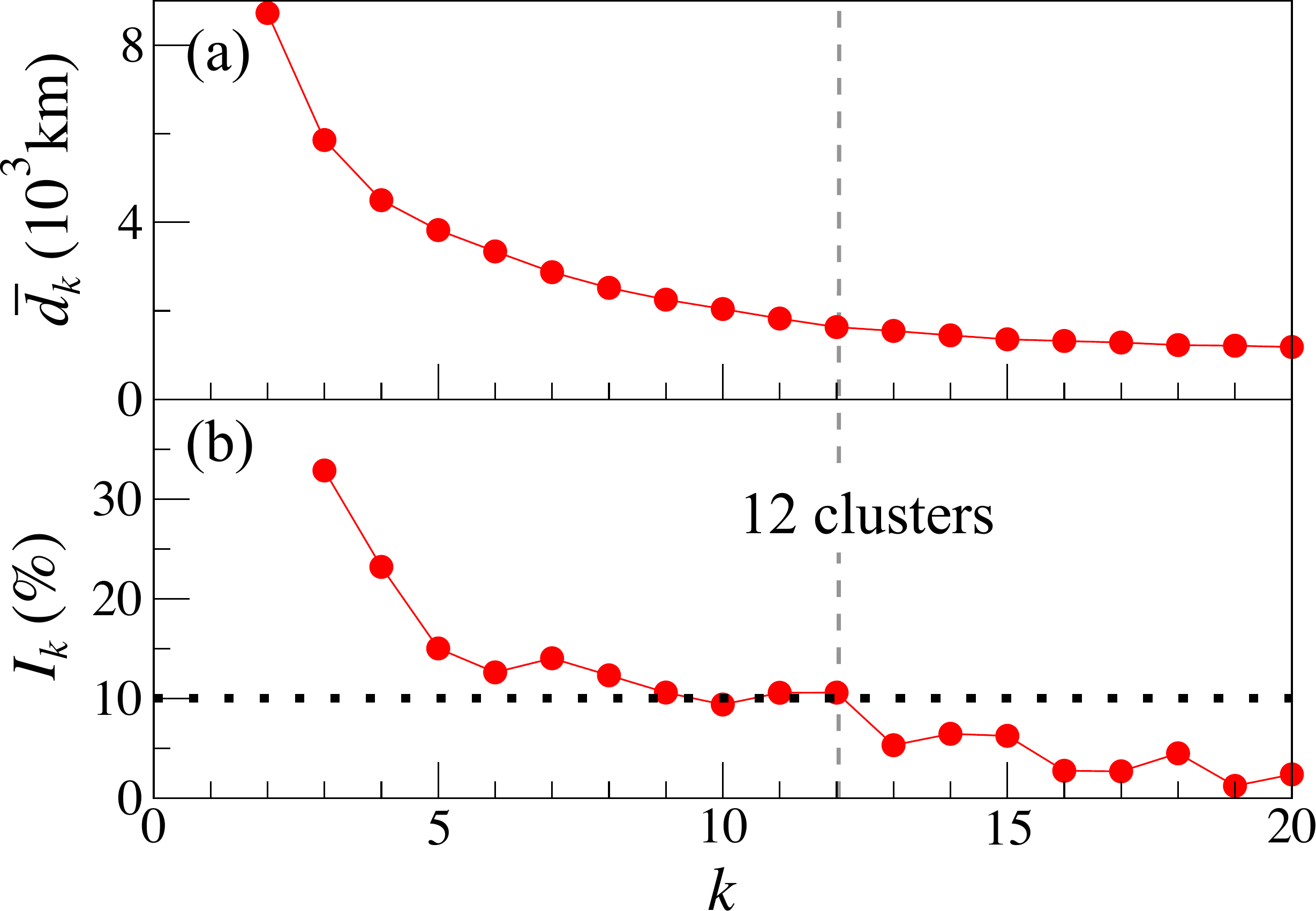}
\caption{Determining the optimal number of clusters $k^*$.
Panel (a) shows that the root-mean-square distance $\bar{d}_k$ 
from an attack location to the center of its assigned cluster 
decreases as the number of clusters $k$ increases. In Panel (b)
we compute the relative change $I_k$ when the number of clusters
increases from $k$ to $k+1$. When $k > 12$, $I_k$
drops noticeably from $\gtrsim10 \%$ to $\lesssim 5 \%$; we thus identify
$k^* = 12$ as the optimal number of clusters.}
\label{FIG:SI_INERTIA}
\end{figure}

We examined the spatial distribution of attacks 
in the AQ and ISIS classes defined above through geographic clustering.
We used the $k$-means algorithm which assigns each event, in this 
case an AQ or ISIS attack, to one of $k$ clusters by iteratively updating the
centers of these clusters and minimizing the root-mean-square distance $\bar{d}_k$ between
the event location and its assigned cluster center \cite{LLO82}.
The number of clusters $k$ is a prescribed parameter for the algorithm;
$\bar{d}_k$ decreases as $k$ increases, as shown in Fig.\,\ref{FIG:SI_INERTIA}(a),
and is exactly zero when
$k$ equals the total number of events, since in this case the location of 
each event becomes its own cluster center. 
Although the goal of minimizing $\bar{d}_k$ favors the choice of a large $k$, decreases in 
$\bar{d}_k$ as $k$ increases may become negligible beyond a threshold value $k^*$, indicating that new clusters are
not distinguishable from old ones. 
The optimal $k^*$ is often determined by 
plotting ${\bar d}_k$ as a function of $k$ and identifying the value of $k$ beyond which 
it begins to plateau. Fig.\,\ref{FIG:SI_INERTIA}(b) illustrates how we
quantitatively identify $k^*$. We first define
$I_k \equiv \vert \bar{d}_{k+1} - \bar{d}_k  \vert / \bar{d}_k$ as 
the relative change in $\bar{d}_k$ as the number of clusters is increased
from $k$ to $k+1$ and plot $I_k$ versus $k$. We find that $I_k$ decreases from
about $10 \%$ to about $5 \%$ when $k$ increases from $12$ to $13$ and stays
well below $10 \%$ as $k$ further increases. We thus set $k^* = 12$ as the
optimal number of clusters. When $k = 13$, the Somalia cluster is split in two
due to its elongated geographic shape; 
since all attacks in Somalia have been attributed to AQ-affiliated al-Shabaab, the
cluster should be unique, confirming that $k^* =12$ is the optimal $k$ value.
We set a threshold of at least 100 post-2014 data points per cluster to justify further analysis.
According to this definition, among the twelve identified above only six contain
enough data to be used for our near-repeat analysis. They are Iraq, Somalia, Syria, Yemen, Nigeria,
and Afghan-Pakistan. Of the others, five clusters have too few attacks for any statistically significant analysis; 
they are Mali, Algeria/E.U., Bangladesh, the Philippines, and the U.S. The Libya cluster has relatively more data points 
than the previous five, but the attacks are sparsely distributed on a vast area,
yielding an insufficient number of closely separated pairs for us to analyze.

As seen in Fig.\,1 of the main text, the twelve clusters are roughly limited by geo-political boundaries
and few attacks occur near the borders; this allows us to name each cluster by the country where the majority of attacks 
took place. It is important to note that our $k$-means clustering uses only
attack-location as input and that we did not pre-impose that clusters be limited
by national borders, rather their geographic extent emerged naturally. 
To verify the robustness of our clustering, we varied the random number sequence of the $k$-means algorithm
seeding the initial cluster centroids, and found that the spatial extent of the clusters remained consistent across runs.  
While most clusters coincide with a single country, a few contain parts of a neighboring one 
due to porous frontiers, or shared political and/or historical traits, showing that diffusion of terrorism 
across boundaries may be possible under specific geographical conditions \cite{KAR14}. 
For example, Lebanon is clustered with Syria due to spill over-effects of the Syrian civil war into Lebanon facilitated by fluid boundaries
between the two. Similarly, since AQ used Somalia as a base to launch attacks against Kenya, the two are part of the same cluster. 
The Nigeria cluster includes a small area between neighboring Chad, Niger and Cameroon
where occasionally Boko Haram has spilled over due to border porosity. 
Afghanistan and Pakistan are in the same cluster due to militant groups residing 
in the tribal corridor between the two, particularly in North and South Waziristan:
the Tehrik-i-Taliban Pakistan (TTP) and the Khorasan Chapter of ISIS.  
Finally, attacks in Algeria, Tunisia and the E.U. coalesce into one cluster 
due to the small, uncoordinated, number of attacks on European soil and in Tunisia which are geographically closest to the denser
ones in Algeria.
For the most part however attacks are mostly confined within nation states. This may be due
to increased military security at the border, or because borders coincide with 
natural barriers such as mountain ranges, deserts or rivers where terrorist events would cause fewer victims, 
elicit less interest from the press, and garner less attention from the population. 
Another possible reason is that terror groups prioritize attacks on ``soft'' targets where large numbers of civilians aggregate and these
are mostly located in major cities, typically in the interior.  
Furthermore, terrorists may prefer to act in familiar settings, responding
to local sources of discontent and grievances \cite{GOL11, LAF10, KIL11}.  Also note that while AQ and ISIS both aim for the supremacy of Islamic values, 
they also establish themselves in territories with pre-existing militant groups
that carry different sources of discontent, instabilities, antipathies, that have existed for much longer periods than the advent of either, 
so that regionality is to be expected on some level. 
Finally, AQ is flexible and operates as a geographically 
diffused network of semi-autonomous cells, allowing for regionality to emerge by design.  
ISIS is more centralized, yet as it conquered or accepted groups that pledged allegiance to it, it divided its territory into provinces, 
factoring in pre-existing conflicts and geographical constraints.  Indeed, the provinces often coincide with the countries (or subnational units) they are based in. 
So although AQ and ISIS are transnational groups, their activities on the ground are tied to the local discourse and remain
clustered mostly in well defined areas. This is also verified by the location of attacks
and origin of their perpetrator groups being always in the same cluster, except for a handful of exceptions. 
For example, the Nusra Front and the Free Syrian Armies concentrated
all of their attacks in the Syria cluster, al-Shabaab stays confined to the Somalia cluster, 
Boko Haram's sphere of action is the Nigeria cluster. 
Finally, note that we investigated possible near-reaction activities over 20km and at the borders between Syria-Lebanon, Somalia-Kenya,
Afghanistan-Pakistan, and across the countries that comprise the Nigeria cluster, and found scant data due to violent attacks being executed near major cities as described earlier. 

\newpage

\section*{AQ and ISIS in the context of global terror}

\begin{figure}[t]
\centering
\includegraphics[width=0.6\textwidth]{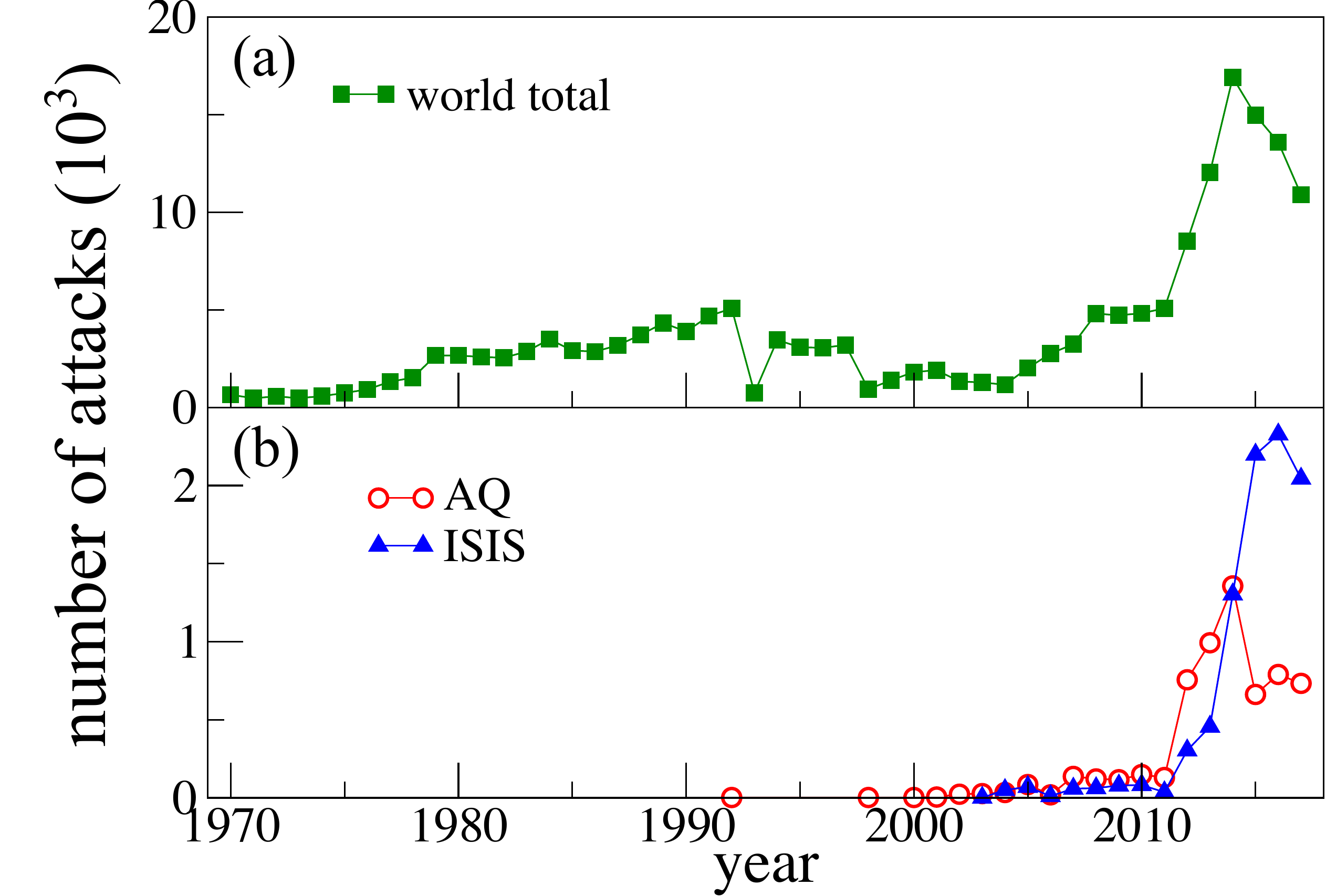}
\caption{Yearly number of terrorist incidents from 1970 to 2017. Panel (a) shows 
the total number of attacks worldwide; panel (b) plots
the number of attacks attributed to AQ and ISIS including their respective official affiliates.
Terrorist activities peaked in the mid 2010s and have been declining since. 
}
\label{FIG:SI_IYEAR}
\end{figure}

Figs.\,\ref{FIG:SI_IYEAR}(a) and (b)  show the total number
of terrorist attacks worldwide and those attributed to AQ
and ISIS respectively between 1970 and 2017. The AQ and ISIS world percentages
shown in Fig.\,2(a) of the main text are derived from these values.
The total number of global attacks increased from
1970 to the early 1990s during the so-called New-Left terrorist 
wave (including attacks from the Red Brigades in Italy, 
RAF in Germany, FARC in Colombia, and Shining Path in Peru to name a few) 
which faltered after the dissolution of the Soviet Union.
A subsequent terrorist wave emerged from religious conflicts 
in the 2000s and climaxed in the 2010s. 
AQ and ISIS follow, and are in part responsible for, the religious
terrorist wave: AQ was engaged in sporadic attacks during the 1990s,
over the next decade it intensified its activities as the ISIS predecessors 
emerged. As a result, the number of attacks attributed to AQ and ISIS grew steadily during the 
2000s and peaked in the early 2010s. 
The global number of terrorist attacks has been decreasing since 2015, and contributions from
AQ/ISIS have stagnated since. The GTD does not report data beyond Dec 31, 2017.
\begin{figure}[h]
\centering
\includegraphics[width=0.5\textwidth]{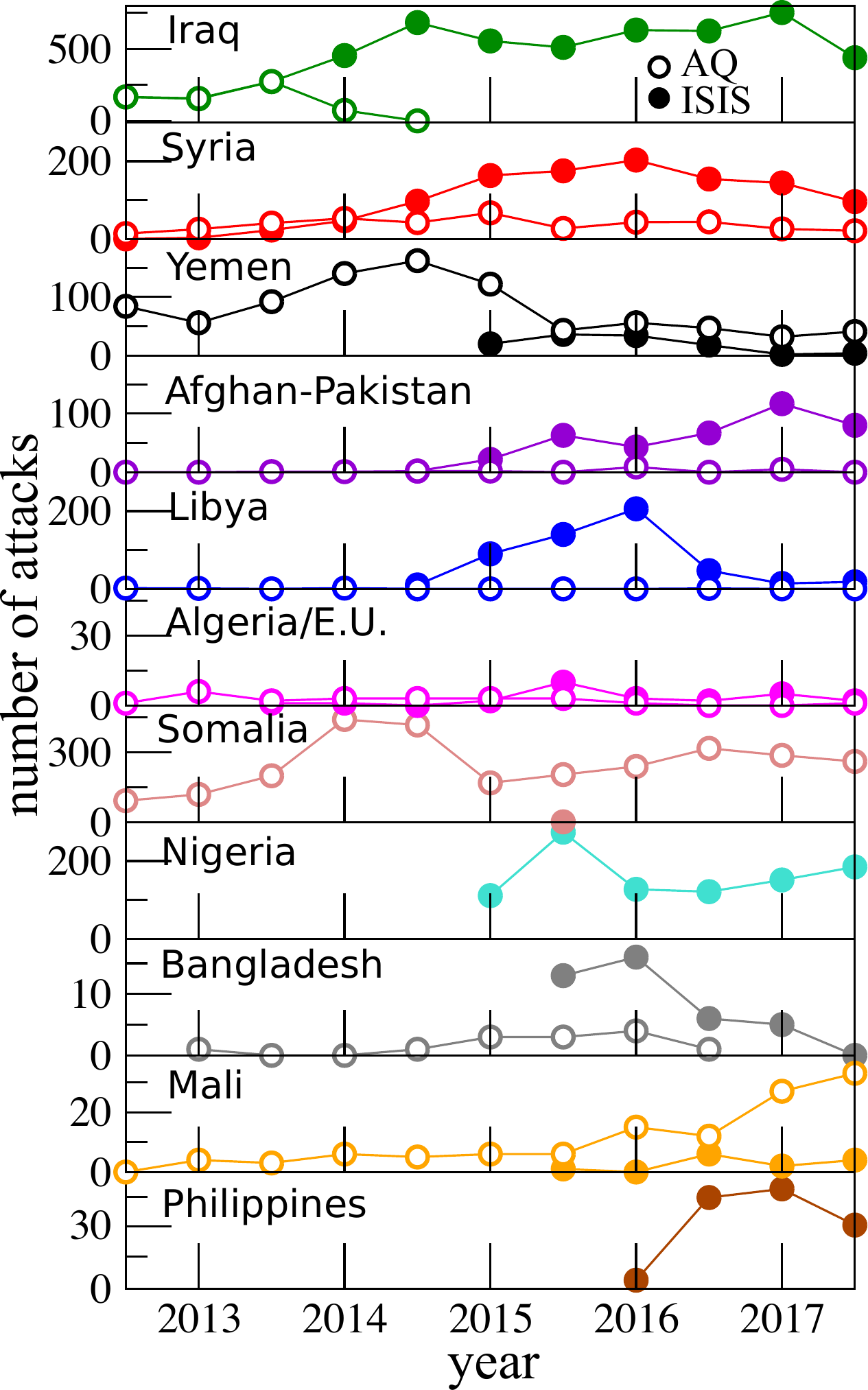}
\caption{Number of AQ and ISIS attacks per half year
in each geographic cluster after 2012. AQ attacks
are represented by open circles; ISIS attacks
are denoted by solid circles. Significant temporal
overlap between AQ and ISIS activity is observed in the Syria, Yemen, and
Bangladesh clusters after the 2014 AQ/ISIS rift.
Clusters are arranged from top to bottom
in the order of their distance from Iraq. The first
data point for the ISIS attacks shifts towards later times in 
the lower panels, revealing that ISIS spread 
from Iraq first to neighboring countries and later to more
distant regions. This figure is complementary to Fig.\,3(b) of the main text.
AQ shows no such pattern and ceased to operate in Iraq post-2014.
}
\label{FIG:SI_TEMPORAL}
\end{figure}

\begin{figure}[h]
\centering
\includegraphics[width=0.41\textwidth]{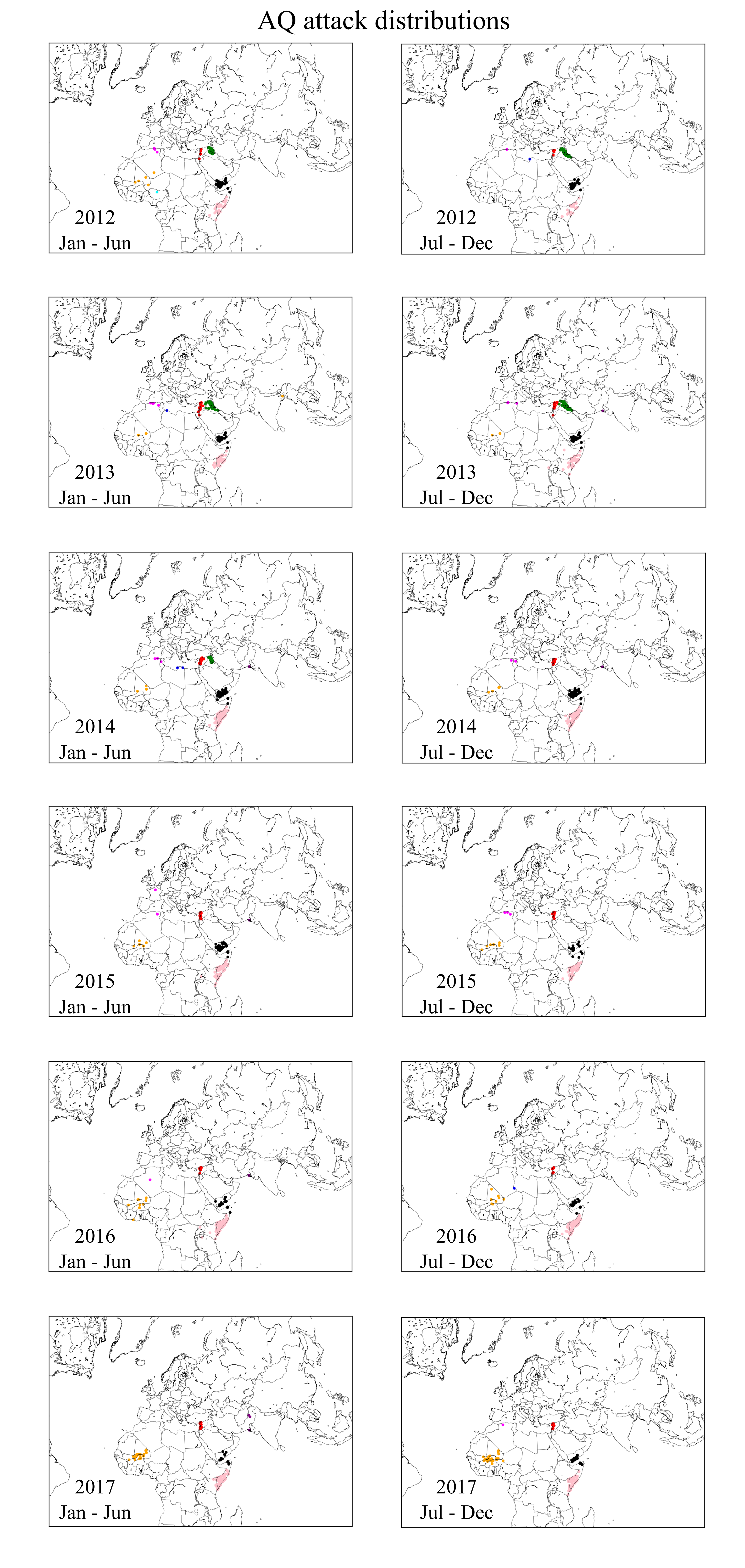}
\includegraphics[width=0.41\textwidth]{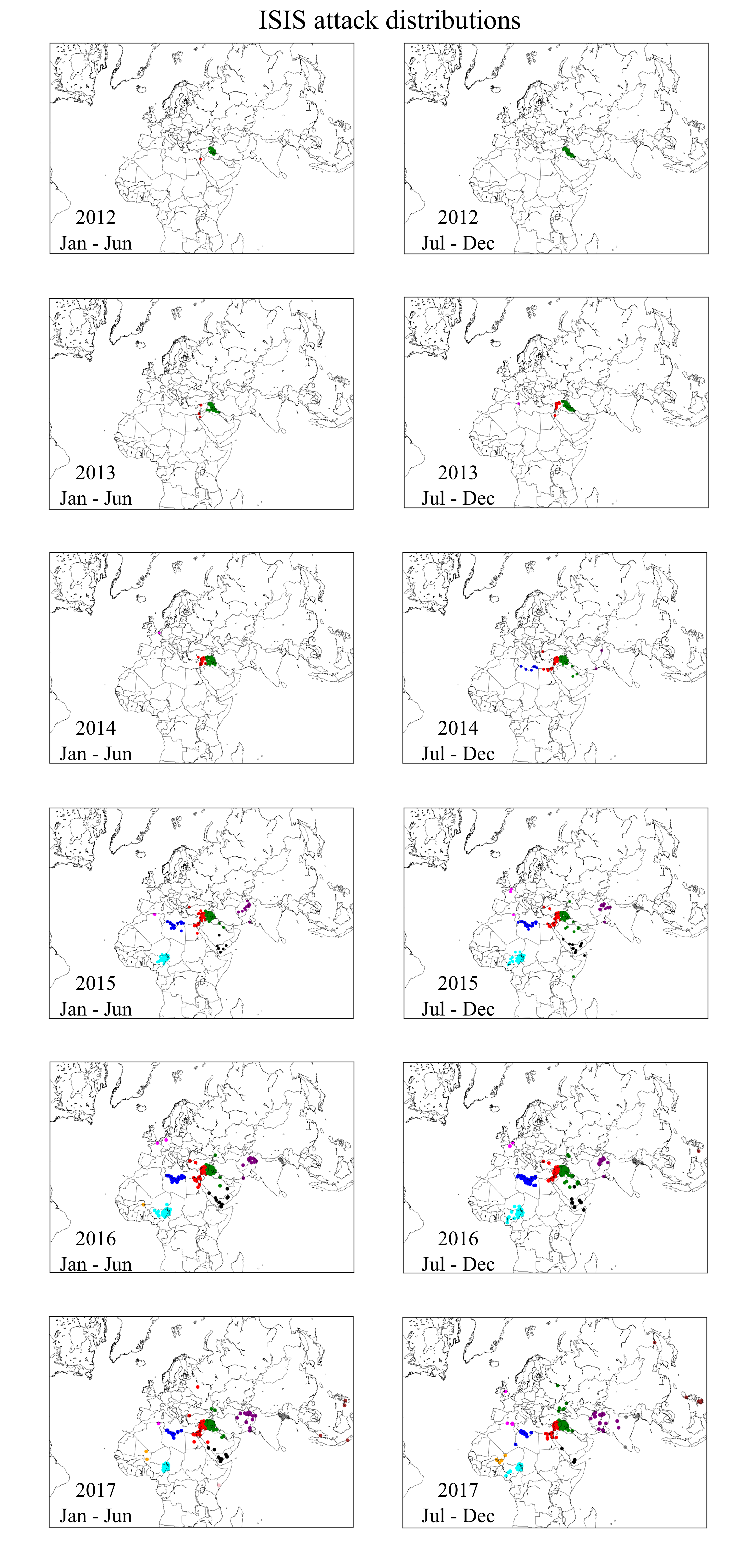}
\caption{Distributions of AQ and ISIS attacks in 
the twelve AQ/ISIS clusters plotted every six months from 2012 to
2017. AQ attacks shift around the globe, emerging
and disappearing from one period to the next. In contrast,
ISIS has maintained a strong presence in Iraq since 2012,
and its attacks have spread out first to Syria in 2014,
to other Middle-East neighboring countries in 2015,
and then to more remote regions after 2016.
}
\label{FIG:SI_AQIS1217}
\end{figure}

\section*{Temporal overlap of AQ and ISIS activities in each cluster}

Fig.\,\ref{FIG:SI_TEMPORAL} shows AQ and ISIS activities within each cluster
as a function of time by binning terrorist incidents in six month intervals starting from 2012.
What emerges is that AQ attacks mostly cease in Iraq 
after it disavowed ISIS in 2014. Significant AQ/ISIS overlap
is observed in Syria and Yemen. In Syria, this overlap is due to ISIS 
expanding into the country in 2013, right before the 2014 AQ/ISIS rift  \cite{BAC17} .
In Yemen, it is due to the establishment of a new ISIS province in 2015. 
The twelve panels are arranged in increasing
order of distance from Iraq. The first data points for ISIS
emerge at later times in the lower panels, 
geographically further from Iraq, revealing a 
spreading terror ``wave'' for ISIS. A complementary visualization is
offered in Fig.\,3(b) of the main text, 
as well as in right panels of Fig.\,\ref{FIG:SI_AQIS1217}, 
where attacks are geo-spatially mapped every six months beginning in 2012. 
AQ activities in contrast manifest more random spatiotemporal variation, 
as can be seen from the left panels of Fig.\,\ref{FIG:SI_AQIS1217}.
The different spatio-temporal patterns characterizing AQ and ISIS  
may be a reflection of their different organizational structures, as AQ may be described as a
decentralized ``dune-like'' network of various cells, while ISIS 
follows a centralized hierarchy. These structural differences are related to their distinct 
goals and approaches \cite{ZEL14}. AQ's priority is to eliminate Western influence from the
Middle East; terrorist attacks are tools to weaken their perceived
oppressors. ISIS's preferred action is to expand the territories it
controls, attacking unfriendly communities and non-Sunni Muslims.
As a result AQ has developed a large decentralized network of collaborative relationships including with the
Taliban in Afghanistan and Pakistan, with al-Shabaab in Somalia, and with Boko
Haram in Nigeria.  In these regions AQ rarely operates on its own, but provides logistic and financial support to
local groups, largely without friction. Very few attacks can thus be directly attributed to AQ, and they are not enough
to be statistically analyzed.  Conversely ISIS's strategy has long been to either
subordinate existing groups, as attempted with Boko Haram in Nigeria, 
or to establish its own terrorist province antagonizing local groups,
as done in Yemen, Afghanistan and Pakistan.

\newpage
\section*{The Kullback-Leibler Divergence and the near-repeat 20 km threshold}

\subsection*{Converting geographic coordinates to distances}
\begin{figure}[t]
\centering
\includegraphics[width=0.3\textwidth]{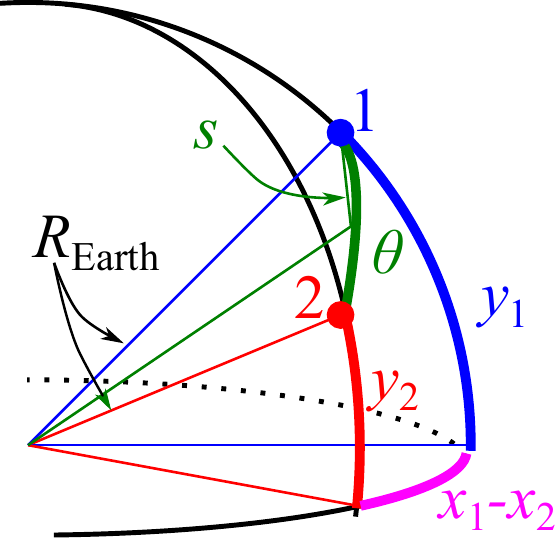}
\caption{Evaluation of the distance $L$ between two geographic locations $i=1,2$ schematically 
represented by the blue and red dots. Each location is associated to $(x_i, y_i)$ coordinates 
where $x_i$ indicates longitude and $y_i$ latitude, for $i=1,2$. 
The evaluation of $L = R_{\rm {Earth}} \theta$ follows Eqs.\,\ref{dist}-\ref{distf} and uses basic trigonometry.
}
\label{FIG:SI_DISTANCE}
\end{figure}

For near-repeat analysis, we need to first compute the distance between 
pairs of GTD events, which are recorded as
longitudinal and 
latitudinal coordinates $(x_i, y_i)$ for all $i$ entries. 
By approximating the Earth as a perfect sphere, as illustrated in Fig.\,\ref{FIG:SI_DISTANCE},
we determine the distance $L$ between two events as follows

\begin{equation}
\label{dist}
   L = R_{\rm Earth} \theta,
\end{equation}

\noindent
where the radius of the Earth is $R_{\rm Earth}$ = 6373 km and where
the radiant angle between the two events is given by 

\begin{equation}
\label{angle}
    \theta = 2 \tan^{-1} \sqrt{\frac{s^2}{R^2_\mathrm{Earth}-s^2}}.
\end{equation}

\noindent
The length of the segment $s$  between the two events that appears
in Eq.\,\ref{angle} is given by 

\begin{eqnarray}
\label{distf}
s &=& R_{\rm Earth} \left[ \sin^2 \left( \frac{y_1 - y_2}{2} \right)
  + \cos \left( y_1 \right) \cos \left( y_2 \right)
 \sin^2 \left( \frac{x_1 - x_2}{2} \right) 
\right]^\frac{1}{2}.
\end{eqnarray}

\noindent
where $x_i$ represents longitude and $y_i$ represents latitude
for the $i = 1,2$ locations.

\newpage
\subsection*{The Kullback-Leibler divergence}
We then utilize the Kullback-Leibler divergence
(KLD) to quantify discrepancies between the observed near-repeat latent time distribution and  
the random event hypothesis (REH) distribution given by $P_d(t)$ in Eq.\,1 of the main text
for pairs of events separated by a distance $L \le d$. 
The KLD is 
defined as the Shannon entropy of the data $\{ (t_i, \hat{p}_i) \}$ 
relative to the REH $\{ (t_i, p_i = P_d (t_i)) \}$

\begin{equation}
\label{kld}
   {\rm KLD} = \sum_i E_i \equiv \sum_i \hat{p}_i \ln \frac{\hat{p}_i}{p_i}.
\end{equation}

\begin{figure}[t]
\centering
\includegraphics[width=0.5\linewidth]{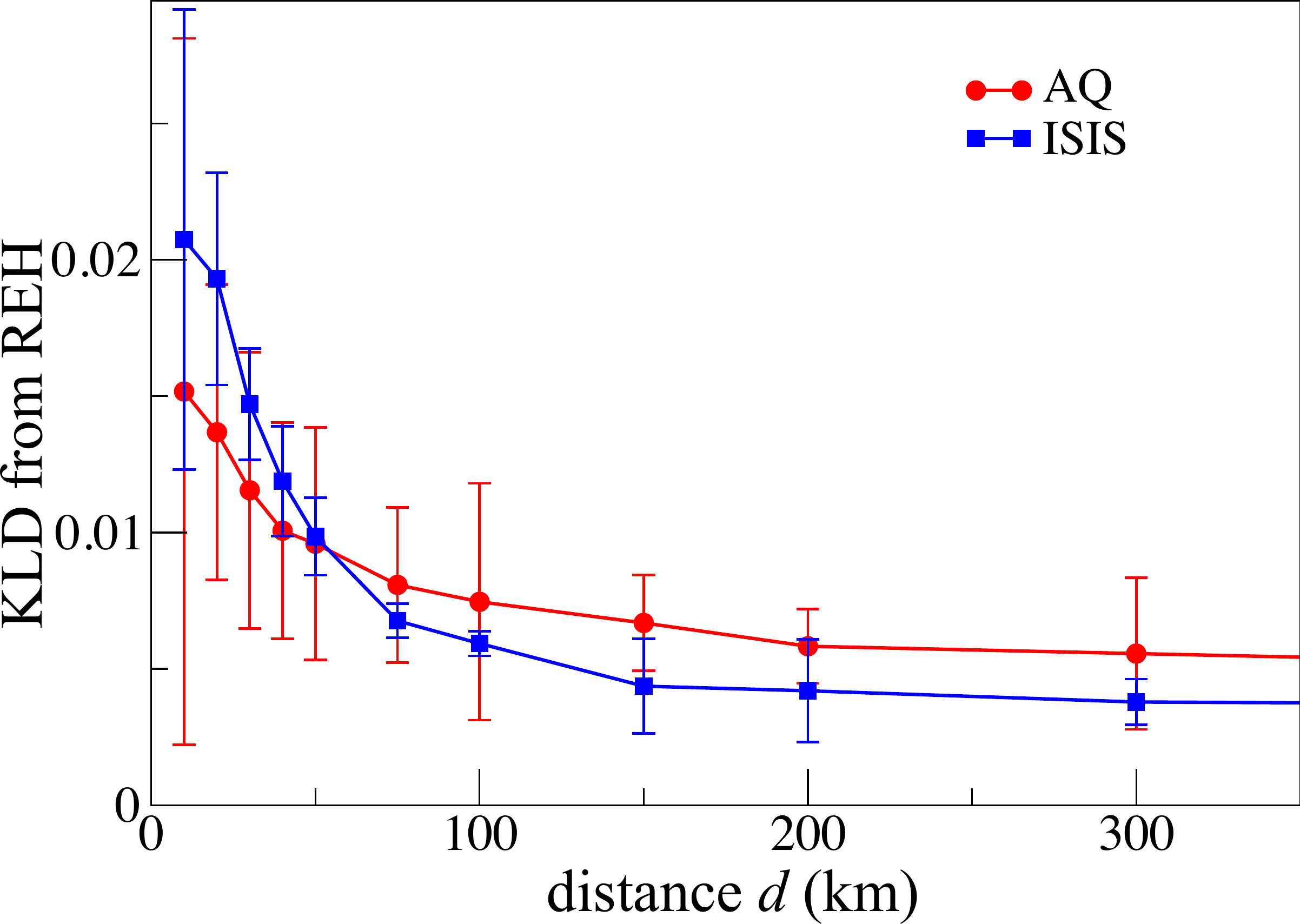}
\caption{The Kullback-Leibler divergence (KLD) as function of the distance
threshold $d$ for AQ and ISIS post-2014. The KLD measures the deviation
of observed near-repeat latent time distributions from the random-event hypothesis (REH).
Near-repeat latent time distributions are constructed for pairs of 
AQ$\to$AQ and ISIS$\to$ISIS attacks separated by a geographical distance limited by the
upper threshold $d$.
Ten periods of $4w = 176$ weeks with different start dates are sampled over
the $204$ weeks between Feb 2, 2014 and Dec 31, 2017. 
Resulting averages and error bars are shown for various values of $d$. 
The curves indicate that a distance threshold of $d=20$km
provides the most distinguishable ranges of KLD values from
those with thresholds larger than 100km. 
Overall the KLD decreases for both AQ and ISIS as $d$ increases.
At short distances, ISIS exhibits larger deviation from the REH
than AQ, revealing relatively stronger near-repeat
tendencies.  At long distances, however, the ISIS data deviates from the REH 
less than the AQ data, suggesting possible long-distance coordination of AQ activities
through its global network.  
}
\label{FIG:SI_KLD}
\end{figure}

\noindent
In Eq.\,\ref{kld} $E_i = \hat{p}_i \ln ({\hat{p}_i}/{p_i})$ is the 
$i^{\rm th}$ data point contribution to the KLD, 
which quantifies how much $\hat{p}_i$ deviates from $p_i$.
The prefactor $\hat{p}_i$ places greater weight on events of
higher probability and reduces the contribution 
of fluctuations associated to rare events.  The 
KLD values depend on $d$, the maximum distance between two events 
 for which the time lag $t_i$ can be calculated. In Fig.\,\ref{FIG:SI_KLD} we plot
the KLD as a function of $d$ for AQ$\to$AQ and ISIS$\to$ISIS near-repeat 
events post-2014. 
There are $204$ weeks between Feb 2, 2014, the official AQ/ISIS rift date,  
and Dec 31, 2017, the last GTD entry. 
We calculate the KLD on a sample of four consecutive $w=44$ week windows within the above time
frame, and repeat the same procedure ten times by randomly changing the start date 
beyond Feb 2, 2014.  The average KLD value and the error bars
shown in Fig.\,\ref{FIG:SI_KLD} are obtained over the effective 40 samples
of duration $w$. 
For both AQ and ISIS, the KLD decreases as $d$ increases, indicating that the near-repeat tendency is stronger 
for events geographically closer to each other, 
and that the latent time distribution converges to the REH when the underlying events
are sufficiently far.  
In the main text we set $d=20$km as the distance threshold within which to study near-repeat phenomena.
This value of $d$ is represented by the second point on each of the curves in 
Fig.\,\ref{FIG:SI_KLD}. The first point corresponds to a $10$km threshold and has the largest KLD value but also
the largest error due to the fewer pairs of near-repeat events that can be constructed 
under a smaller upper distance limit. 
When $d=20$km instead, the KLD is large enough and the error 
small enough, to distinguish it from KLD values at greater distances, say 
at $100$km. 

We also find it statistically significant that for attacks within $20$km ISIS expresses a 
relatively stronger near-repeat tendency compared to AQ, since ISIS has greater KLD values than AQ
and the two data points reside outside their respective error ranges.
As $d$ increases, the KLD value of AQ decreases at a slower rate than ISIS, 
the gap between them closes, and for distances of 
several hundreds of kilometers the KLD value of AQ exceeds that of ISIS.
This may suggest a certain long-range coordination of AQ attacks which may be 
facilitated by its global network structure. 
We verified that all results are robust to moderate changes in $w$ and to
start dates beyond Feb 2, 2014. 

For near-reaction patterns, we compute the correlation $r$ between the  
$\{ E_i^{\rm A \to B} \}$ and $\{ E_i^{\rm B \to A} \}$ datasets, derived respectively from the
A$\to$B and B$\to$A panels in Fig.\,5 of the main text, where $\{$A,B$\}$ = $\{{\rm AQ, ISIS, L }\}$
and ${\rm A} \neq {\rm B}$.
A positive $r$ indicates heightened attack response probabilities for both the A and B class organizations in response
to each other's attacks; this may be due to collaborative, aligned, or retaliatory copycat events, for example. 
Conversely a negative $r$ implies that the attack likelihood of say, class A quickly increases
in response to attacks by class B, whereas class B delays its response after attacks by class A. 
This asymmetry may be to to structural differences between the two classes, where for example
class B requires more time to evaluate and organize response attacks to class A. 
One of the main findings from the main text is 
that when the three AQ, ISIS, and L classes are all present in a cluster, 
one of the transnational groups (AQ or ISIS) and the local L class
emerge as major players, and are in conflict with each other. The remaining
transnational group (ISIS or AQ) is instead a minor player and tends to align itself
with either of the major ones. These dynamics are reflected 
in the post-2014 near-reaction patterns.  Asymmetric A$\to$B and B$\to$A near-reaction
activity, and a negative $r$ are found to characterize rival organizations, where the  
agile, quick to respond A is the local militia/insurgent
group and B the slower transnational organization (AQ/ISIS). 
Symmetric A$\to$B and B$\to$A near-reaction activity and a positive $r$ is instead 
a hallmark of collaborative/aligned groups.  As we shall describe below, results from pre-2014 data 
are consistent with the post-2014 analysis illustrated above
and rival groups still manifest A$\to$B and B$\to$A near-reaction asymmetry.  
There is no distinction between AQ and ISIS in any geographical cluster 
pre-2014, because either they were either the same organization,
or ISIS had not expanded into the region yet; as a result, no collaborative/aligned relations, and no symmetric
 A$\to$B and B$\to$A near-reaction patterns are identified.
Small $\vert r \vert$ indicates weak correlation; in this case the sign of $r$ becomes
irrelevant since the two classes are essentially not responding to each other.
We use conventional criteria and assume that $r \ge 0.66$ ($r \le -0.66$) indicates strong
correlation (anti-correlation), $\vert r \vert < 0.33$ represents weak correlation,
and $0.33 < r \le 0.66$ ($-0.33 > r \ge -0.66$) is indicative of intermediate correlation
(anti-correlation).

\section*{Outbidding}

\begin{figure}[t]
\centering
\includegraphics[width=0.9\linewidth]{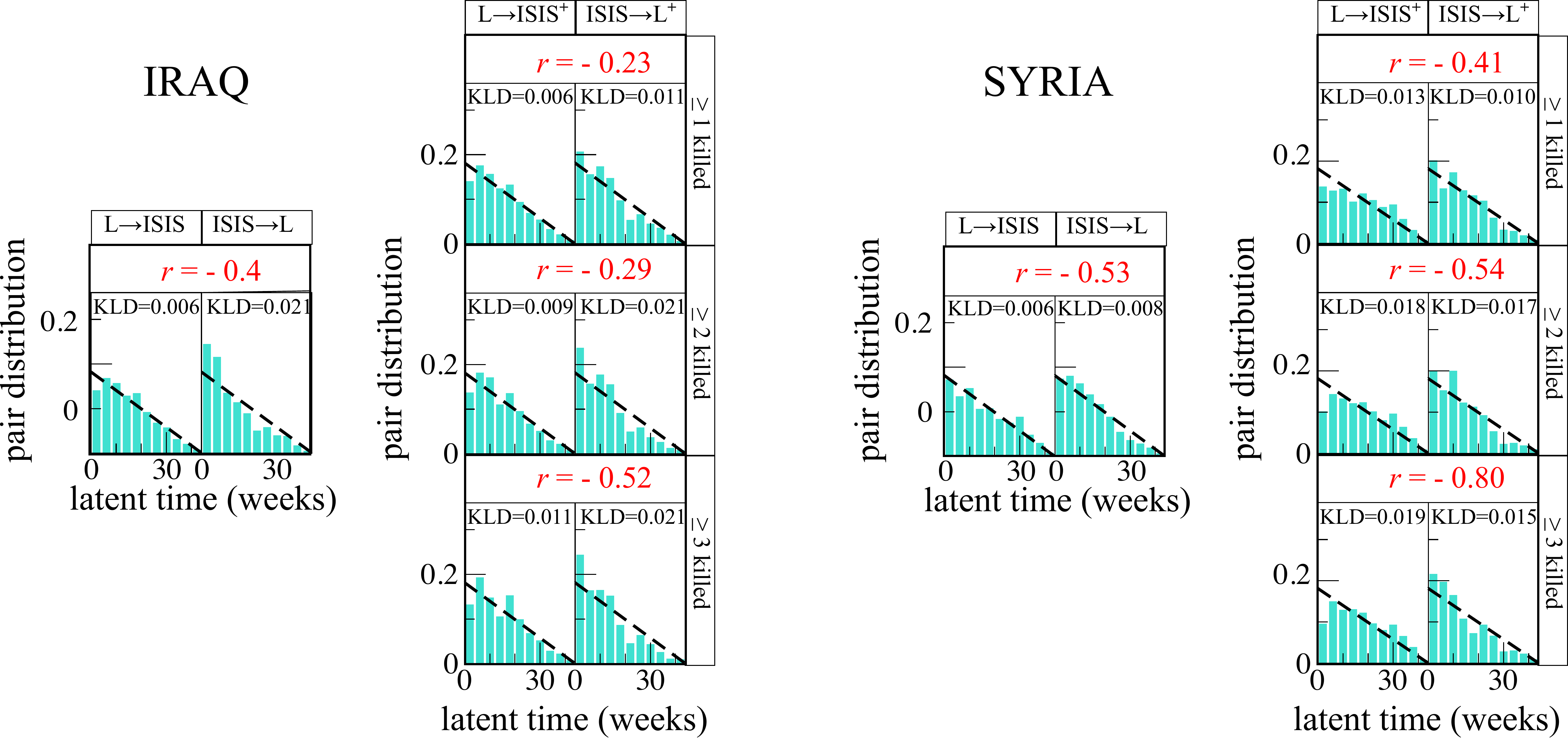}
\caption{IRAQ (left): Near-reaction panels L$\to$ ISIS (ISIS $\to$ L) for attacks within 20km of each other; 
IRAQ (right): Near-reaction panels L $\to$ ISIS$^{+}$ (ISIS $\to$ L$^{+}$)
with at least one, two or three casualties in the response attack. Note the L$\to$ ISIS$^+$ (ISIS $\to$ L$^+$) and the L$\to$ ISIS
(ISIS $\to $ L) panels carry similar features: no essential changes can be detected. The nimble L class responds quickly to ISIS attacks, 
whereas ISIS will require longer decisional and/or organizational time to respond, regardless of lethality. 
The same trends are seen in the Syria panels. Incremental outbidding, mutually lethal outbidding, or (in the case of Iraq) 
suicide attack near-repeat reveal no sign of outbidding.}
\label{FIG:OUTBID}
\end{figure}
\begin{figure}[h!]
\centering
\vspace{0.5cm}
\includegraphics[width=0.6\linewidth]{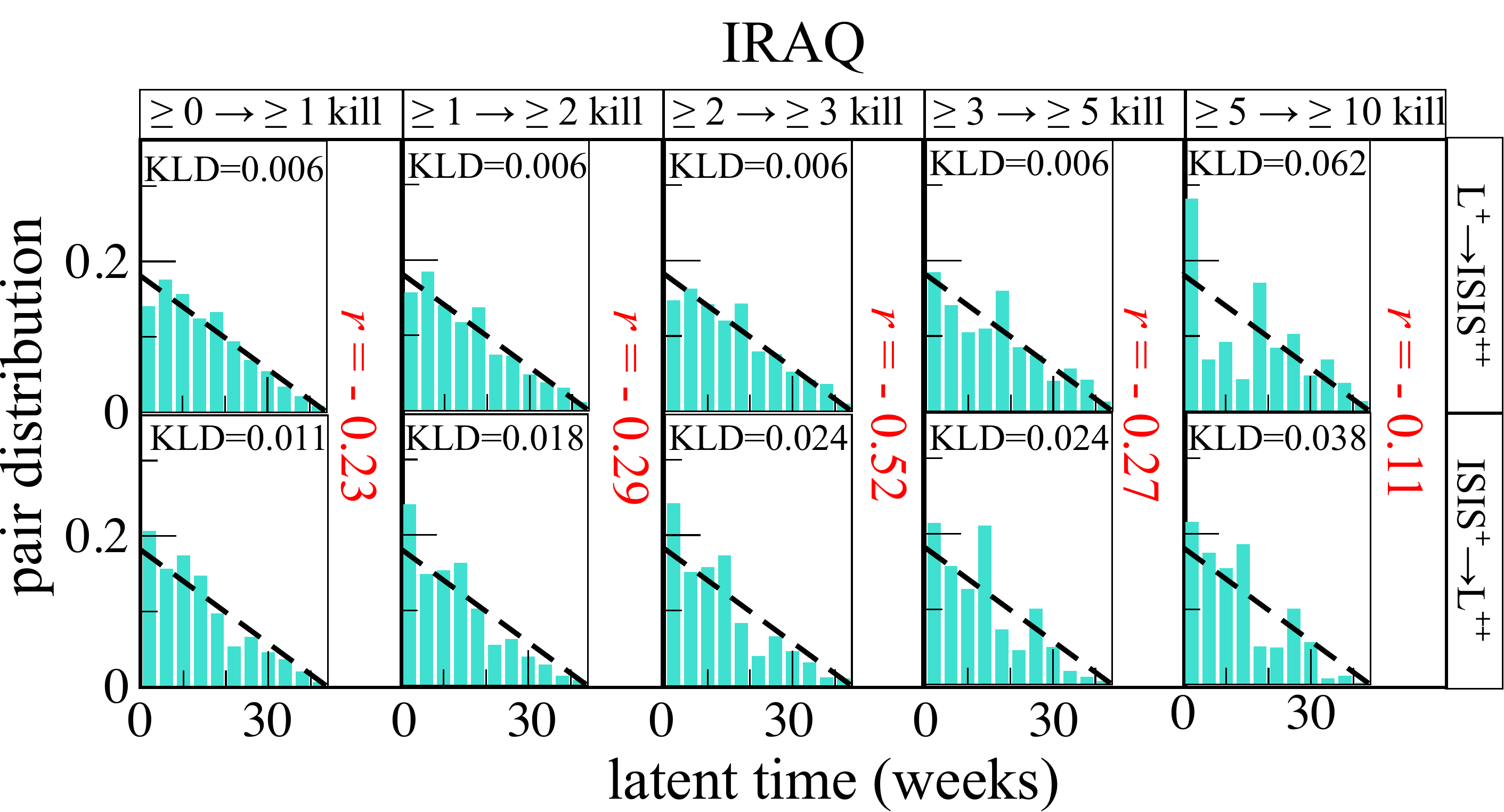}
\includegraphics[width=0.6\linewidth]{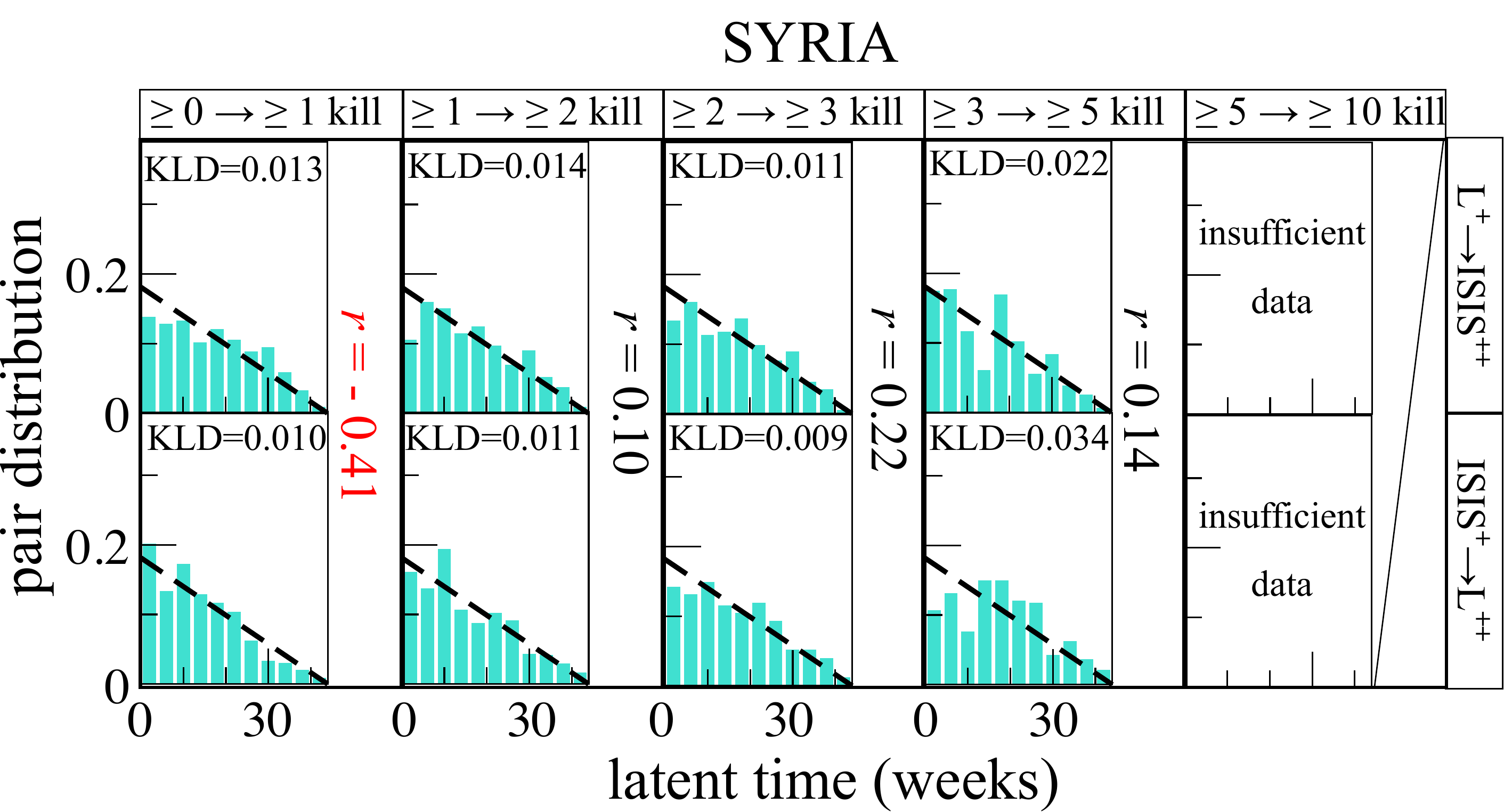}
\caption{Iraq (left): Near-reaction panels L$^+$ $\to$ ISIS$^{++}$ and ISIS$^{+}$ $\to$ L$^{++}$, where deadly 
attacks trigger deadlier attacks within 20km of each other.
Stratifying on the basis of the number of casualties does not yield any novel trend, apart from the already noted
ISIS/L response asymmetry as they react to each other's attacks. The same inconclusive trends are seen in the Syria panels.}
\label{FIG:OUTBID2}
\end{figure}

Outbidding is the process of competitive escalation among two (or more) separate but related groups who operate 
in the same geographical area and who orchestrate increasingly violent attacks to 
outshine the other(s) \cite{BLO04, KYD06}. The underlying assumption is that greater violence
shows greater commitment and/or capability and can help garner support from the population. 
This mechanism was first proposed in Bloom, 2004 to explain 
escalating suicide bombings in Palestine by Hamas and Fatah, 
vying to be championed by the citizenry. The First and Second Intifadas are sometimes cited as
examples of this practice.  
Similar phenomena have been described in scenarios that include nationalist, left and
right-wing political violence \cite{POR13}, or within ethnic conflicts \cite{HOR85}. At times,
caveats have been included such that outbidding will emerge only if communities are supportive
of suicide bombings, or if religiously or nationalistically motivated \cite{NEM14}. 
Other authors have questioned the principle of
outbidding leading to escalation of suicide terrorism and, in general, of terrorist acts of any type. 
For example worldwide data from 1970 to 2004 was analyzed in \cite{FIN12}. Apart from Israel, 
little support is found for outbidding, leading the authors to warn of the dangers
of overgeneralizing from a limited set of cases.  Others yet have criticized describing even the conflict in Palestine
as an outbidding process and have excluded it from occurring in Iraq, at least prior to the advent of
ISIS \cite{BRY08, AYE08}. A frequent objection to outbidding as a way to bolster support from the local population 
is that if this were the case, there would be fewer attacks with unknown offenders as terror groups
would better advertise their actions. For the GTD data we analyzed about 46$\%$ of the total
number of relevant attacks was due to unknown perpetrators. 
We examined the possibility of competitive escalation between the major terrorist groups in each cluster where applicable. 
In particular we consider the Iraq and Syria clusters, where the major players are ISIS and the L class. Instead of filtering for the number
of suicide attacks, we analyze near-reaction patterns filtering for the number of casualties, 
as data is more plentiful.  

The first scenario considered, shown in Fig.\,\ref{FIG:OUTBID} is that of general outbidding, where any type of attack is followed
by a lethal attack from rival groups with \textit{at least} one, two or three casualties.
These panels are denoted as L $\to$ ISIS$^{+}$ (ISIS $\to$ L $^+$) where the presence of 
casualties is represented by the $+$ sign.  For comparison we also replot the L $\to$ ISIS (ISIS $\to$ L) panels
from the main text. As can be seen, the duration of the near-reaction time window,
and the general shape of the attack pair distribution does not change much when lethality is taken into account 
compared  to when it is.
The only trend that arises is the same observed in the  L $\to$ ISIS (ISIS $\to$ L) panels: 
the more nimble L class will more swiftly respond to ISIS attacks, to generate more support or attention, 
while ISIS will delay its response. The same findings are found in the Syria shown in the right hand
side of Figure \ref{FIG:OUTBID}.  
Another possibility is that of an incremental outbidding scenario, 
where a lethal attack with X casualties is followed by another with at least Y $>$ X casualties
(L$^{+}$ $\to$ ISIS$^{++}$). Various choices for X,Y are shown in Fig.\,\ref{FIG:OUTBID2}; however
just as above no novel trend can be identified compared to the L $\to$ ISIS (ISIS $\to$ L) panels. 
We also considered mutually lethal outbidding, where any deadly attack is followed by any other deadly attack,
L$^+$ $\to$ ISIS $^{+}$. Finally, for Iraq we could also strictly select for suicide attacks,
whereas this not possible in Syria due to insufficient data. No novel patterns were found in any of these
cases. The same outcome emerged from pairing major and minor players in all other clusters. 

Finally, we examined provoked outbidding where (B $\to$ A)$\to$ B attack sequences are compared
to general A $\to$ B sequences.  In the (B $\to$ A$)\to$ B chain of events, 
B strikes first and the A response is constrained to be within 20km and 4 weeks, defining a first
near-reaction response. We then take this subset of A attacks, which can be thought of as provoked by B,
and study how the B class responds to them, defining a second near-reaction response.
Our goal is to determine whether the second B near-reaction response in the (B $\to$ A$)\to$ B
sequence differs from the general B response in the unprovoked A $\to$ B sequence. 
Indeed, if outbidding were at play, we would expect escalation to be manifest 
in B's response to provoked A attacks, (B $\to$ A)$\to$ B,  through larger deviations from the REH than 
B's response to general, unprovoked A attacks, A$\to$ B.

\begin{figure}[t]
\centering
\vspace{0.5cm}
\includegraphics[width=0.5\linewidth]{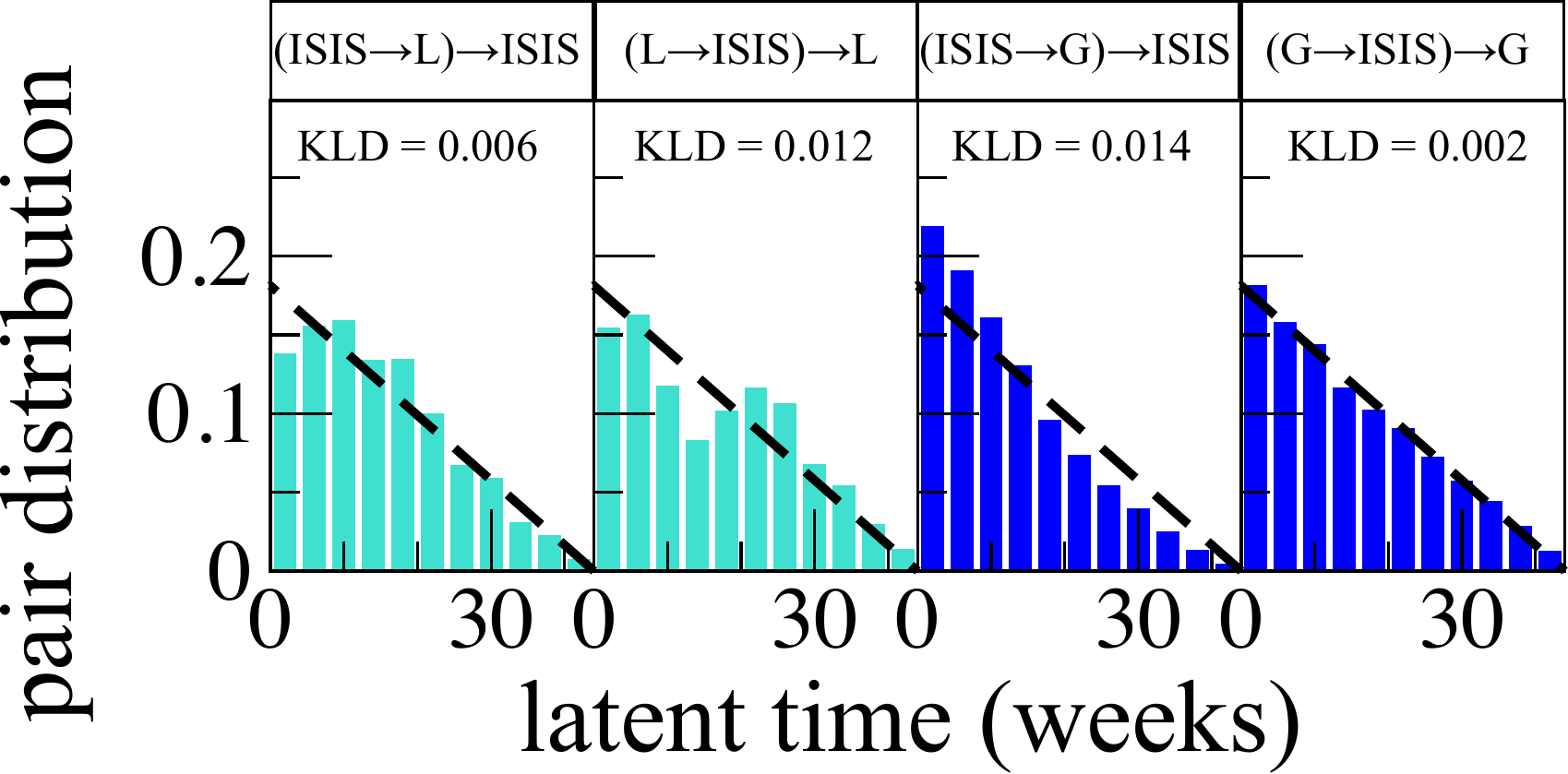}
\caption{Provoked attack near-reaction panels in Iraq. ISIS responses are not enhanced 
when responding to provoked attacks by the L class, compared to when responding
to general, unprovoked L class events. This can be seen by contrasting the (ISIS $\to $L) $\to $ISIS panel here 
with the L $\to$ ISIS one in Fig.\,\ref{FIG:OUTBID}. Similarly no provoked outbidding is observed for
L responding to ISIS if one compares the (L $\to $ISIS) $\to $L panel here 
with the ISIS $\to$ L one in Fig.\,\ref{FIG:OUTBID}.  No provoked outbidding is seen either between ISIS and the Iraqi government
forces, which is evident by comparing the (ISIS $\to$ G$)\to$ ISIS panel here with the G $\to$ ISIS one in Fig.4(c1) of the main text
or by comparing the (G $\to$ ISIS$)\to$ G panel here with the ISIS $\to$ G panel in Fig.4(c1) of the main text.}
\label{FIG:OUTBID3}
\end{figure}

Only the Iraq cluster contains sufficient data for a meaningful analysis. 
Among the 4575 post-2014 ISIS attacks here, 105 were provoked by L according to the criteria defined above; 
among the 996 L attacks, 151 were provoked by ISIS. 
Conversely, among the 4575 post-2014 ISIS attacks in Iraq, 1330 were provoked by G; 
among the 2219 G interventions, 1728 were provoked by ISIS. The above ISIS/G data shows considerable interplay between terrorist
activity and the Iraqi government. 
Provoked near-reaction panels in the Iraq cluster are shown in Fig.\,\ref{FIG:OUTBID3}: we find 
no evidence of provoked outbidding either between L and ISIS, nor between G and ISIS. 
The panel for (ISIS $\to $L) $\to $ISIS shows 
in fact slightly suppressed near-reaction compared to the general L $\to $ISIS case
displayed in Fig.\,\ref{FIG:OUTBID}; similar trends are observed for the (L $\to $ISIS) $\to $L case. 
We also investigated provoked outbidding involving the Iraqi government and ISIS. 
The (ISIS $\to$ G$)\to$ ISIS panels show elevated near reaction as ISIS responds to provoked
government operations, however the response is not noticeably different or stronger than ISIS responding to general G
events, as can be seen in the G $\to$ ISIS panel Fig.\,4(c1) of the main text.
Both show elevated ISIS response likelihood within the first 4 months and the same degree of deviation from REH
(KLD=0.014 in both provoked and unprovoked government operations). No outbidding is observed in the (G $\to$ ISIS$)\to$ G events either: 
G responds to ISIS retaliations indifferently, indicated by the near-REH latent time distribution. 
The same analysis is not possible for Syria due to lack of sufficient data. 

Our work thus can be placed within the literature warning of the application of outbidding as a general
theory: near-response and near-repeat trends do not seem to greatly change when escalation is included. 
Perhaps, apart from specific cases such as Palestine and its unique Hamas-Fatah rivalry, 
terrorist groups operate in the same way whether or not attacks involve
casualties; whether casualties arise may also be largely out of their control.  This is also
confirmed by recent studies which explicitly exclude competitive escalation
between AQ and ISIS \cite{HAM17b}.

\section*{Government intervention}

The interplay between counterterrorism efforts and terrorist attacks \cite{BLA07} is studied in the Iraq, Somalia and Afghan-Pakistan clusters, since they are the only 
ones where sufficient data is available. As mentioned in the main text,
we utilize the Uppsala Conflict Data Program (UCDP) dyadic dataset \cite{PET18} to do this. 
The data is collected from global media sources and listed in terms of dyads engaged in armed conflict; 
players may include state actors. A dyad is considered to be in conflict if both sides adopt incompatible positions 
that lead to more than 25 casualties within a year, in which case all related activities are recorded. 
We cross-list this data with terrorist events from the
GTD to create near-reaction diagrams. 
In Iraq, between 2014-2017 the UCDP lists 2230 incidents involving the Iraqi government directed at ISIS;
the GTD lists 4362 attacks for ISIS and 996 L-class events in the same time interval.
For Somalia, we study the effects of counterterrorism activity between Feb 9, 2012, the day that
al-Shabaab officially joined the AQ, and Dec 31, 2017. Coincidentally,
the Federal Government of Somalia was established on August 20, 2012. 
The UCDP lists 1847 events by the Somali government against al-Shabaab
while the GTD lists 2850 al-Shabaab attacks within the same timeframe.  
The Taliban government was overthrown from Afghanistan in 2001, following the US-led invasion that took place
after the 9/11 attacks. For the Afghan-Pakistan cluster thus a natural timeframe for terrorist and counterterrorist near-reaction studies
would be the post-2001 period. Note that the 2014 AQ/ISIS rift is unlikely to play a relevant role here
since the Taliban were never officially part of either AQ or ISIS. Before 2003 however, the UCDP logs only one incident, whereas the GTD contains only 8 Taliban attacks.
This is most likely due to the relatively quiet period following the Taliban's defeat:  
immediately after their 2001 capitulation it underwent an internal reorganization 
and launched its first insurgency against the Afghan government 
in 2003. We thus conduct our analysis of the Afghan-Pakistan cluster within the Jan 1, 2003 -- Dec 31, 2017 time interval.
Here, the GTD lists 9406 Taliban attacks and 394 ISIS attacks between 2003-2017; 
the UCDP lists 21030 instances of anti-terrorist intervention. 
Data for Syria is not available, as the UCDP only recently began translating its polygon 
dyad system into geographic coordinates; data for Nigeria is not sufficient for a meaningful analysis.
Due to the volatile and complex civil war in Yemen, there are severe ambiguities in identifying the legitimate government.
Our finding of enhanced near-reaction activity between government and terrorist activity, with long term deviations from the REH in all clusters for which data was available, 
suggests that state-sponsored military actions may yield increased levels of violence, at least in their immediate aftermath. 
Although we consider domestic state actors as compiled by the UCDP (the Iraqi, Somali and Afghan governments)
our findings are consistent with a recent study on foreign military intervention conducted in 122 countries 
between 1970 and 2015 where foreign counterterrorism action was similarly associated to a short-term increase in terror incidents
\cite{PIA19}.

\section*{Pre-2014 data}

\begin{figure}[t]
\centering
\includegraphics[width=0.5\linewidth]{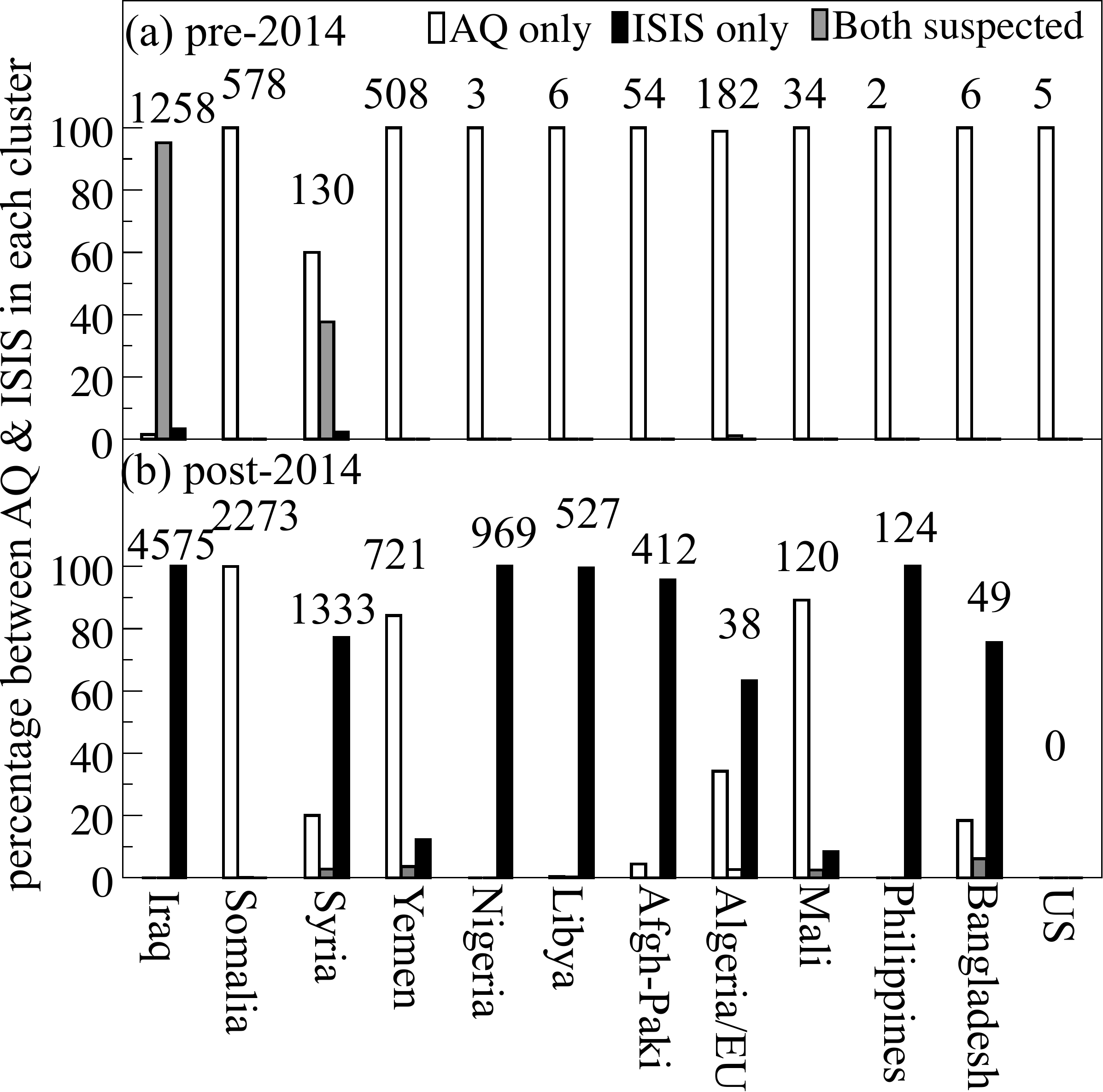}
\caption{Percentages of attacks by AQ and ISIS, including their respective affiliates,
in each of the twelve clusters. Panel (a) shows pre-2014 data, when ISIS was mostly confined to
Iraq and Syria as an AQ affiliate. Panel (b) shows post-2014 data, after ISIS 
began spreading to other regions independently from AQ. Post-2014, AQ's activities concentrate in Somalia, Yemen, Mali;
in all other clusters where activities are reported ISIS is the most active. 
The total number of AQ and ISIS attacks per cluster and
period is labeled above the bar graphs. Total numbers increased from pre-2014 to
post-2014 in Iraq, Somalia, Syria, Yemen, Afghan-Pakistan, and Mali; the Nigeria,
Libya, Philippines, and Bangladesh clusters went from nearly no attacks to tens or
hundreds of attacks. Conversely, post-2014 attacks in the Algeria/E.U. cluster
decreased; there have been no been direct AQ/ISIS attacks in the U.S. post-2014.
}
\label{FIG:SI_EACH_CLUSTER}
\end{figure}

In Fig.\,3(a) of the main text, we show how attacks are distributed 
between AQ and ISIS in each cluster during the time period 2001-2017.
Figs.\,\ref{FIG:SI_EACH_CLUSTER}(a) and (b) compare
these percentages before and after the 2014 AQ/ISIS rift.
Since ISIS was largely confined to Iraq and Syria at the time, 
the pre-2014 attacks in all other clusters are almost exclusively due to AQ as shown in
Fig.\,\ref{FIG:SI_EACH_CLUSTER}(a), except for a very small fraction ($1 \%$) in the Algeria/E.U. 
cluster.  In Iraq, $95 \%$ of pre-2014 attacks were perpetrated
by ISIS and its various
predecessors as part of AQ's global network, $3 \%$ were by
the ISIS predecessor JTJ, which was never officially an AQ affiliate,
and $2 \%$ by other small AQ cells in the region. 
In Syria, $60 \%$ of attacks were perpetrated by AQ-affiliated JN; 
ISIS as an affiliate of AQ was responsible for $38 \%$ of attacks,
mostly occurring during 2013 when it began expanding into Syria. 
The remaining $2 \%$ were earlier 2002 attacks in Jordan
by JTJ.  After the AQ/ISIS 2014 rift ISIS overtook AQ as the most attack-prolific terrorist group
in most clusters, including Iraq, Syria, Nigeria, Libya, Afghan-Pakistan,
Algeria/E.U., the Philippines, and Bangladesh,
as shown in Fig.\,\ref{FIG:SI_EACH_CLUSTER}(b).
Post-2014 AQ maintained its dominance only in
Somalia, Yemen, and Mali.

\begin{figure}[t]
\centering
\includegraphics[width=0.9\linewidth]{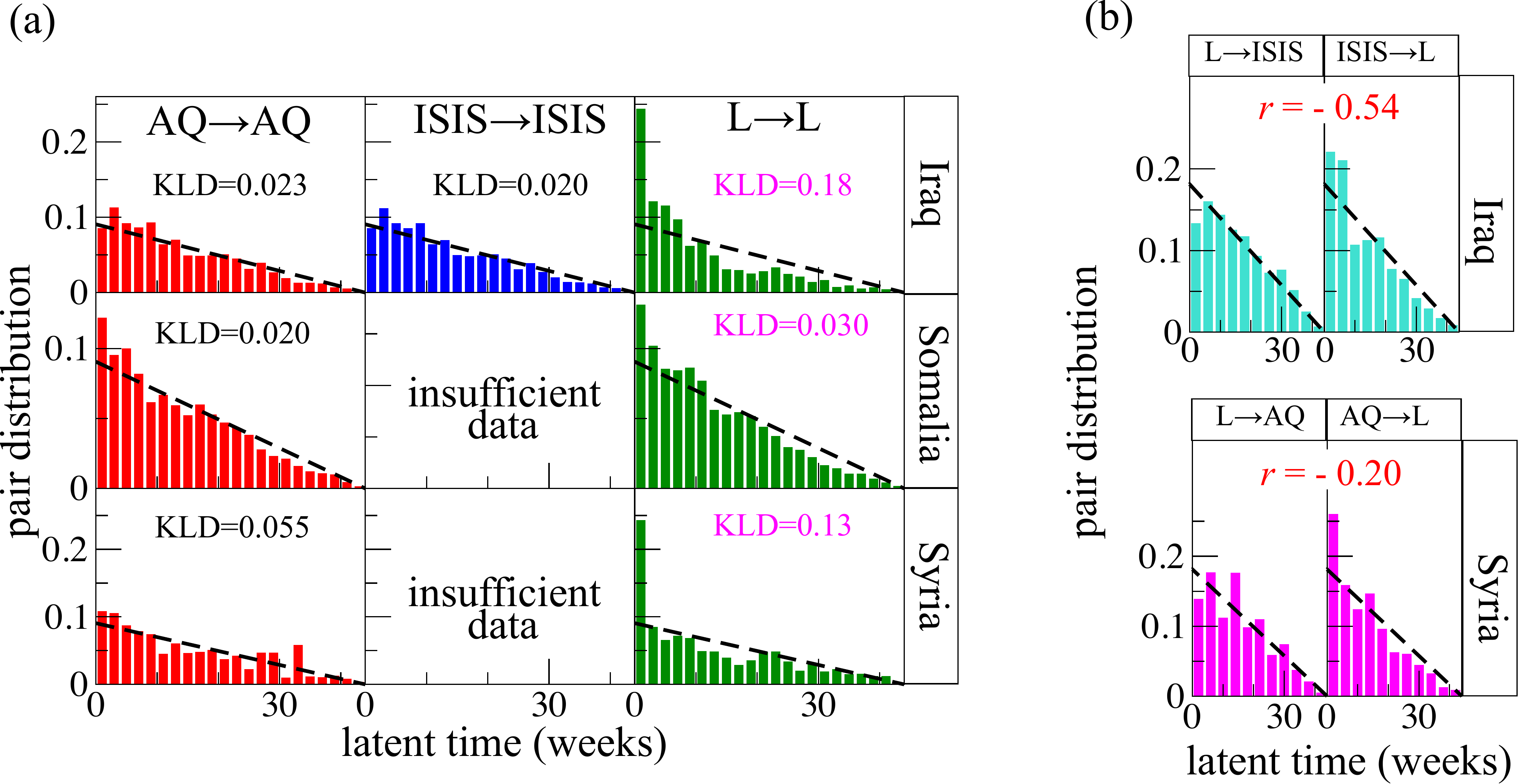}
\caption{(a) Pre-2014 near-repeat patterns of AQ, ISIS, and L classes in Iraq, Somalia, and
Syria. In Iraq and Syria AQ/ISIS were major players compared to 
the local L class. In Somalia the AQ$\to$AQ panel refers to al-Shabaab post-2012 (and pre-2014)
after it officially joined AQ, while the L$\to$L panel refers to al-Shabaab pre-2012, while it still operated
as an independent group. The KLD quantifying the deviation of the observed near-repeat data
from the REH is evaluated for each panel; 
the largest per cluster is highlighted in magenta.
Note that the L class exhibits the greatest deviation
from the REH in both Iraq and Syria, consistent with the hypothesis that minor players manifest larger
near-repeat likelihoods compared to major ones. No such distinction is possible in Somalia
since the only group operating here is  al-Shabaab, either as an AQ affiliate or as an independent entity.
(b) Pre-2014 near-reaction patterns ISIS$\to$L and L$\to$ISIS in Iraq and AQ$\to$L and L$\to$AQ in Syria. 
Both cases result in asymmetric latent time distributions and negative $r$
between AQ/ISIS$\to$L and L$\to$AQ/ISIS. The negative correlation between
ISIS and L in Iraq is moderately significant, and is consistent with the hypothesis of 
adversary relations between the two classes. The correlation between AQ and L in Syria is weak.
}
\label{FIG:SI_NRA_PRE2014}
\end{figure}

Figs.\,\ref{FIG:SI_NRA_PRE2014}(a) and (b) show the pre-2014 near-repeat and near-reaction 
latent time distributions for AQ, ISIS, and L class attacks separated by less than $20$km 
in the Iraq, Somalia, and Syria clusters. The others do not contain sufficient
pre-2014 data for a significant analysis. For example, the
Yemen and Algeria/E.U. clusters display numerous attacks 
that were sparsely distributed over large areas, leading to few near-repeat/reaction 
pairs of events. In contrast, the pre-2014 attacks in Syria
mostly concentrated within two years (2012-2013) over the relatively smaller area
of northern and eastern Syria. 
Fig.\,\ref{FIG:SI_NRA_PRE2014}(a) shows that the minor classes,
defined as those with fewer number of combatants, 
exhibit more prominent near-repeat tendencies than major ones, 
in agreement with results shown in Fig.\,4 of the main text. 
Fig.\,\ref{FIG:SI_NRA_PRE2014}(b) reveals asymmetric 
near-reaction patterns for rival groups so that local militias/insurgent groups 
from the L class respond promptly to AQ/ISIS attacks, whereas
the latter show delayed reactions. This is also consistent with results shown in Fig.\,5 in the main text.
In pre-2014 Iraq, $95 \%$ of AQ/ISIS attacks were attributed to both since
ISIS was considered an AQ affiliate from 2004 to 2014.
As a result, the AQ$\to$AQ and ISIS$\to$ISIS near-repeat 
patterns in Fig.\,\ref{FIG:SI_NRA_PRE2014}(a) are almost identical, 
with KLD = $0.023$ and $0.020$ respectively. 
AQ and ISIS also both show slightly elevated near-repeat likelihood within the first eight weeks
of an initial attack compared to the REH; in contrast, the L class shows a 
much higher near-repeat probability within the first eight weeks after an initial attack, if compared to both
AQ and ISIS, with KLD = $0.18$. 
Most L class attacks in pre-2014 Iraq were attributed to
small groups that executed less than five attacks, in addition to
non-specific perpetrators, such as Sunni/Shia/Muslim extremists, 
pro-government/Baathist extremists, tribesmen, and separatists.
The numerous small terrorist groups most likely reflect the changing sociopolitical 
Iraqi scenario, as the initial insurgency against the U.S.--led
coalition became an insurgency against the Shia Muslim--led Iraqi government. 
The large number of non-specific perpetrators may be due to limited coverage from the 
Western press, and possibly because the turbulent and rapidly changing
terrorism landscape was largely unstructured or unknown to the outside 
world during this period. A case in point is the especially
volatile year 2006.  L class groups associated with more than 10 attacks pre-2014 include 
known ISIS allies such as Ansar al-Islam and Ansar al-Sunna, as well as ISIS opponents, 
such as PKK, the Peace Companies (also known as the Mahdi Army), and Asa'ib Ahl al-Haqq. 
The great number of small groups and non-specific perpetrators raises uncertainties as to 
whether the pre-2014 L class in Iraq acted as an ally or an opponent of ISIS, or even
as to whether it can be considered a unique block.
The general understanding however is that local insurgents mainly expressed
anti-ISIS sentiment, as ISIS's ideological intolerance often exacerbated
sectarian conflicts. ISIS 
experienced hostility even within its own Sunni-Muslim community,
as exemplified by the Anbar Awakening movement between 2005 and 2008, 
when local Sunni militias and tribes rose against ISIS in northern Iraq. 
This conjecture is corroborated by the asymmetric near-reaction patterns between ISIS
and local L class groups in Fig.\,\ref{FIG:SI_NRA_PRE2014}(b), with a moderately significant
negative correlation $r = -0.54$. This implies that L class groups 
responded promptly to ISIS attacks, whereas ISIS delayed its reactions after 
L class attacks, which is consistent 
with the post-2014 analysis given in the main text. 
Since $95 \%$ of AQ and ISIS attacks overlapped in pre-2014 Iraq, 
the AQ versus L near-reaction plots are essentially identical to the ISIS versus L ones
and thus not shown. Similarly, AQ$\to$ISIS and ISIS$\to$AQ near-reaction
patterns are essentially 
the same as AQ$\to$AQ and ISIS$\to$ISIS near-repeat patterns
and are also omitted.

The Somalia cluster is dominated by al-Shabaab, which was officially recognized
as an AQ affiliate in 2012, whereas ISIS is not present in this region. 
The AQ$\to$AQ and L$\to$L panels shown Fig.\,\ref{FIG:SI_NRA_PRE2014}(a) both refer to
near-repeat patterns of al-Shabaab, but at different times. 
Before 2012, as an independent group al-Shabaab exhibits elevated near-repeat likelihood 
over a period of 12 weeks after a first attack, as shown in the L$\to$L panel for Somalia in 
Fig.\,\ref{FIG:SI_NRA_PRE2014}(a). This is longer than the eight-week period shown
in the AQ$\to$AQ panel in the same figure which contains
al-Shabaab 2012-2014 data.
Over the first eight weeks, the near-repeat probability is also slightly higher before 2012 (L$\to$L)  
than after 2012 (AQ$\to$AQ). 
Since the AQ and the L classes here do not overlap in time, 
there is no near-reaction between AQ and L.

In Syria, the majority of pre-2014 attacks took place between
2012-2013 during the Syrian Civil War.  ISIS invaded Syria in April 2013, but
there is not enough data to study ISIS$\to$ ISIS near-repeat patterns. 
Indeed, during most of the pre-2014 era, fighters from ISIS or its precursors
joined AQ-affiliated JN in opposition to other anti-government rebels,
so that in Syria AQ is the major class, followed by the local L class and no significant ISIS 
presence is found.
While both AQ-affiliate JN and L class rebels exhibit heightened near-repeat probability over 
the first four weeks, as shown in Fig.\,\ref{FIG:SI_NRA_PRE2014}(a), 
the near-repeat probability is especially large during the first two weeks 
for the L class, leading to a KLD value of $0.13$, larger than that of AQ-affiliated JN, which is
$0.055$.  
The pre-2014 L class in Syria consists of the same anti-government rebels as in
the post-2014 L class discussed in the main text, including the Free Syrian Army, the PKK, and the Islamic Front.
However, the 2014 AQ/ISIS rift changed the relation between AQ and these L class groups.
AQ came to Syria through its affiliate JN in 2012 which emerged as an opponent
to most of the local rebels. This is reflected in the asymmetric near-reaction AQ$\to$L and L$\to$AQ
patterns shown in Fig.\,\ref{FIG:SI_NRA_PRE2014}(b) marked by a weak negative
correlation $r = -0.2$. The moderate negative correlation value may stem from 
tentative alliances between AQ and local militia groups after the
ISIS invasion of Syria in April 2013 which lead to more symmetric behavior. 
Nonetheless, the overall weakly asymmetric near-reaction pattern
confirms that the response from the local L class groups is enhanced when AQ strikes first, 
whereas reactions from the transnational AQ class are delayed, 
consistent with findings presented in the main text for post-2014 data.
Finally, as discussed in the main text, AQ and local L class antigovernment rebels 
became allies against ISIS in post-2014 Syria, 
leading to symmetric near-reaction AQ$\to$L and L$\to$AQ patterns, 
marked by a positive correlation coefficient $r = 0.61$ as shown in Fig.\,5 of the main text.





\FloatBarrier






\bibliography{aqis}